\documentclass[12pt]{article}
\usepackage{epsfig}

\usepackage[utf8x]{inputenc}
\usepackage{lmodern,textcomp}

\usepackage{latexsym}
\usepackage{indentfirst}
\usepackage{fancyhdr}
\usepackage{amssymb}
\usepackage{amsmath}
\usepackage{amsthm}
\usepackage{amsfonts}
\usepackage{pifont}
\usepackage{cite}
\usepackage{color}
\usepackage{slashed}
\usepackage{graphics}
\usepackage{url}
\usepackage{array}
\usepackage{graphicx}
\usepackage{float}
\usepackage{parskip}
\usepackage{url}

\usepackage[framemethod=TikZ]{mdframed}
\mdfsetup{skipabove=\topskip,skipbelow=\topskip}
\newrobustcmd\ExampleText{%
An \textit{inhomogeneous linear} differential equation has the form
\begin{align}
L[v ] = f,
\end{align}
where $L$ is a linear differential operator, $v$ is the dependent
variable, and $f$ is a given non-zero function of the independent
variables alone.
}
\newcounter{theo}[section]
\newenvironment{theo}[1][]{%
\stepcounter{theo}%
\ifstrempty{#1}%
{\mdfsetup{%
frametitle={%
\tikz[baseline=(current bounding box.east),outer sep=0pt]
\node[anchor=east,rectangle,fill=black!20]
{\strut \thetheo};}}
}%
{\mdfsetup{%
frametitle={%
\tikz[baseline=(current bounding box.east),outer sep=0pt]
\node[anchor=east,rectangle,fill=black!20]
{\strut #1};}}%
}%
\mdfsetup{innertopmargin=10pt,linecolor=black!20,%
linewidth=2pt,topline=true,
frametitleaboveskip=\dimexpr-\ht\strutbox\relax,}
\begin{mdframed}[]\relax%
}{\end{mdframed}}



\usepackage{array,multirow}
\usepackage{hhline}
\usepackage{graphicx}
\usepackage{subcaption}

\begin{document}

\begin{titlepage}

   \vskip 1cm
   \begin{center}
    {\Large\bf Demystifying Gauge Symmetry}
   
   \vskip 0.2  cm
   \vskip 0.5  cm
Jakob Schwichtenberg$^{\,a,}$\footnote{E-mail: \texttt{jakob.schwichtenberg@kit.edu}},
\\[1mm]
   \vskip 0.7cm
 \end{center}

\centerline{$^{a}$ \it  Institut f\"ur Theoretische Teilchenphysik, 
Karlsruhe Institute of Technology,}
\centerline{\it  Engesserstra{\ss}e 7, D-76131 Karlsruhe, Germany} 
\vspace*{1.5cm}

\begin{abstract}
\noindent
Gauge symmetries are often highlighted as a fundamental cornerstone of modern physics. But at the same time, it is commonly emphasized that gauge symmetries are not a fundamental feature of nature but merely redundancies in our description. We argue that this paradoxical situation can be resolved by a proper definition of the relevant notions like "local", "global", "symmetry" and "redundancy". After a short discussion of these notions in the context of a simple toy model, they are defined in general terms. Afterwards, we discuss how these definitions can help to understand how gauge symmetries can be at the same time fundamentally important and purely mathematical redundancies. In this context, we also argue that local gauge symmetry is not the defining feature of a gauge theory since every theory can be rewritten in locally invariant terms. We then discuss what really makes a gauge theory different and why the widespread "gauge argument" is, at most, a useful didactic tool. Finally, we also comment on the origin of gauge symmetries, the common "little group argument" and the notoriously confusing topic of spontaneous gauge symmetry breaking.

\end{abstract}
\end{titlepage}

\section{Introduction}
\label{sec:intro}

Gauge symmetry is one of the most important concepts in modern physics. However, although the original idea is more than a century old, there are ongoing discussion on the meaning, origin and importance of gauge symmetries.\cite{Redhead2003-REDTIO-3,pittphilsci1831,pittphilsci9289,BELOT2003189,Brading2004-BRAAGS,doi:10.1086/687936,Earman2001-EARGM-2,healey2007gauging}

On the one hand, gauge symmetry is a helpful concept that lets us understand the structure of the Standard Model of particle physics.\footnote{For a discussion of the role of gauge symmetry in General Relativity, see Ref.~\cite{Percacci:1984bq}.} Its importance is highlighted regularly in lectures, talks and textbooks. For example:

\begin{itemize}
    \item "\textit{More recently, the principle of local gauge invariance has blossomed into a unifying theme that seems capable of embracing and even synthesizing all the elementary interactions.}" - Chriss Quigg in his popular textbook "Gauge Theories of the Strong, Weak, and Electromagnetic Interactions" \cite{quigg2013gauge}.
    \item "\textit{We now suspect that all fundamental symmetries are local gauge symmetries.}" - David J. Gross in his talk "The triumph and limitations of quantum field theory" \cite{Gross:1997bc}.
    \item "\textit{[A] central pillar of the Standard Model is the idea of gauge symmetry.}" - Frank Wilczek in his talk "Quantum Beauty: Real and Ideal".
    \item "\textit{Gauge invariance is a guiding principle in building the theory of fundamental interactions.}" -  Michele Maggiore in his textbook "A Modern Introduction to Quantum Field Theory" \cite{maggiore2005a}.

\end{itemize}

On the other hand, there is an increasingly large number of prominent physicist who argue that gauge symmetry is not really a fundamental feature of nature but merely a technical redundancy. 

For example:

\begin{itemize}
    \item “\textit{What’s as a misnomer called gauge symmetry, whose beauty is extolled at length in all the textbooks on the subject, is completely garbage. It’s completely content free, there’s nothing to it.}” -  Nima Arkani-Hamed in his talk "Space-Time, Quantum Mechanics and Scattering Amplitudes" \cite{hamedtalk}.
    \item "\textit{Gauge invariance is not physical. It is not observable and is not a symmetry of nature.}" - Matthew Schwartz in his textbook "Quantum Field Theory and the Standard Model" \cite{schwartz2014quantum}.
    \item "\textit{[G]auge symmetries aren’t real symmetries: they are merely redundancies in our description of the system.}" - David Tong in his lectures on "The Quantum Hall Effect" \cite{Tong:2016kpv}.
    \item "\textit{Gauge symmetries are properly to be thought of as not being symmetries at all, but rather redundancies in our description of the system}" - Philip Nelson and Luis Alvarez-Gaume in their paper on "Hamiltonian Interpretation of Anomalies" \cite{Nelson:1984gu}. 
    \item "\textit{[G]auge symmetry is strictly speaking not a symmetry, but a redundancy in description.}" - Anthony Zee in his textbook "Quantum Field Theory in a Nutshell" \cite{zee2010quantum}.
    \cite{wen2004quantum}.
\end{itemize}

Therefore, it is natural to wonder: "\textit{If gauge is only mathematical redundancy, why the common emphasis on the importance of gauge symmetry? Why the idea that this is a major discovery and guiding principle for understanding particle physics?}" \cite{Rovelli:2013fga}

One of the main sources of confusion is that there is no canonical definition of the notion "gauge symmetry". Even worse, "gauge symmetry" and related notions like "local gauge symmetry" are rarely properly defined in papers and books since it is assumed that everyone knows what is meant. 

This root of the problem was already described almost 50 years ago by Andrzej Trautman:

"\textit{Few words have been abused by physicists more than relativity, symmetry, covariance, invariance and gauge or coordinate transformations. These notions used extensively since the advent of the theory of relativity, are hardly ever precisely defined in physical texts. This gives rise to many misunderstandings and controversies.}" \cite{Trautman:1970cy}

The goal of this paper is to unify the various ideas surrounding gauge symmetries and put them into a coherent context.

After a short discussion of symmetries in general in Section~\ref{sec:symmetries}, gauge symmetry and important related notions are discussed in intuitive terms in Section~\ref{sec:gaugesymintuitively} and Section~\ref{sec:gaugetheoryintuitively} using a simple toy model. Afterward, in Section~\ref{sec:gaugesymphysics} the same notions are discussed in more concrete terms in the context of Quantum Mechanics and Electrodynamics. The main result is that global gauge symmetries are real symmetries with observable consequences while local gauge symmetries are redundancies. Among others, these two important notions are discussed in Section~\ref{sec:gaugesymmathematically} in more precise terms. Using these ideas, we discuss in Section~\ref{sec:gaugetheorymathematically} what really makes a theory a gauge theory. In then Section~\ref{sec:gaugetale}, we then argue that the common "gauge argument", which is often used to motivate the structure of gauge theories, does not hold up to closer scrutiny. Finally, in Section~\ref{sec:originofgaugesymmetry} the origin of gauge symmetry and the somewhat popular "little group argument" are discussed.

\section{Symmetries Intuitively}
\label{sec:symmetries}
Before we can discuss gauge symmetry, we have to talk about symmetries in general. 

So first of all, what is a symmetry?

Imagine a friend stands in front of you and holds a perfectly round red ball in her hand. Then you close your eyes, your friend performs a transformation of the ball and afterward you open your eyes again. If she rotates the ball while your eyes are closed, it is impossible for you to find out that she did anything at all. Hence, rotations are symmetries of the ball.

\begin{center}
    \includegraphics[width=0.7\textwidth]{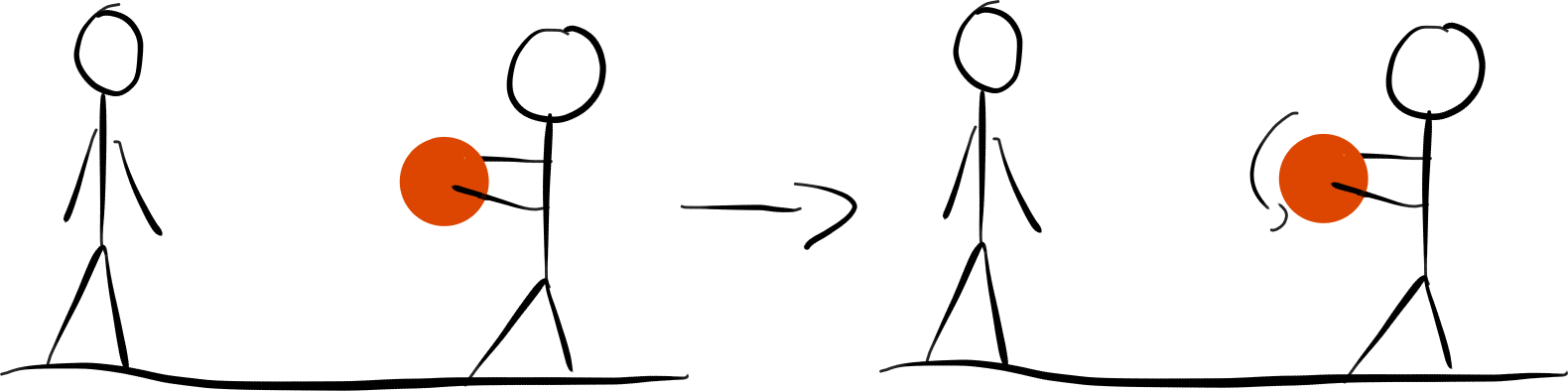}
\end{center}

In contrast, if she holds a cube, only very special rotations can be done without you noticing it. In general, all transformations which, in principle, change something but lead to an indistinguishable result are symmetries. Formulated differently, a symmetry takes us from one state to a different one, which however happens to have the same properties.\footnote{In contrast, a redundancy takes us from on description of a state to another description of the same state. This will be discussed in detail in Section~\ref{sec:activepassive}.}

It's important to take note that with this definition, symmetries are \textit{observable} properties of objects or systems. This is especially important in the context of gauge symmetries because here we need to be careful what transformations are really symmetries.

This point can be confusing at first and to understand it, we need to talk about subsystems and global or local transformations of them.

\subsection{Global vs. Local Symmetries}

First, let's consider a \textbf{global transformation} of a subsystem, say a ship. Global means here that the whole ship is rotated and not just some part of it, which would be a \textbf{local transformation}.

Clearly a global rotation of a ship takes us from one state to a physically different one since we are actively rotating the ship. However, if there is a physicist inside the ship with no possibility to look outside, there is no way for him to find out whether he is in the original or the rotated ship.\footnote{This thought experiment is known as Galileo's ship experiment.}

\begin{center}
    \includegraphics[width=0.9\textwidth]{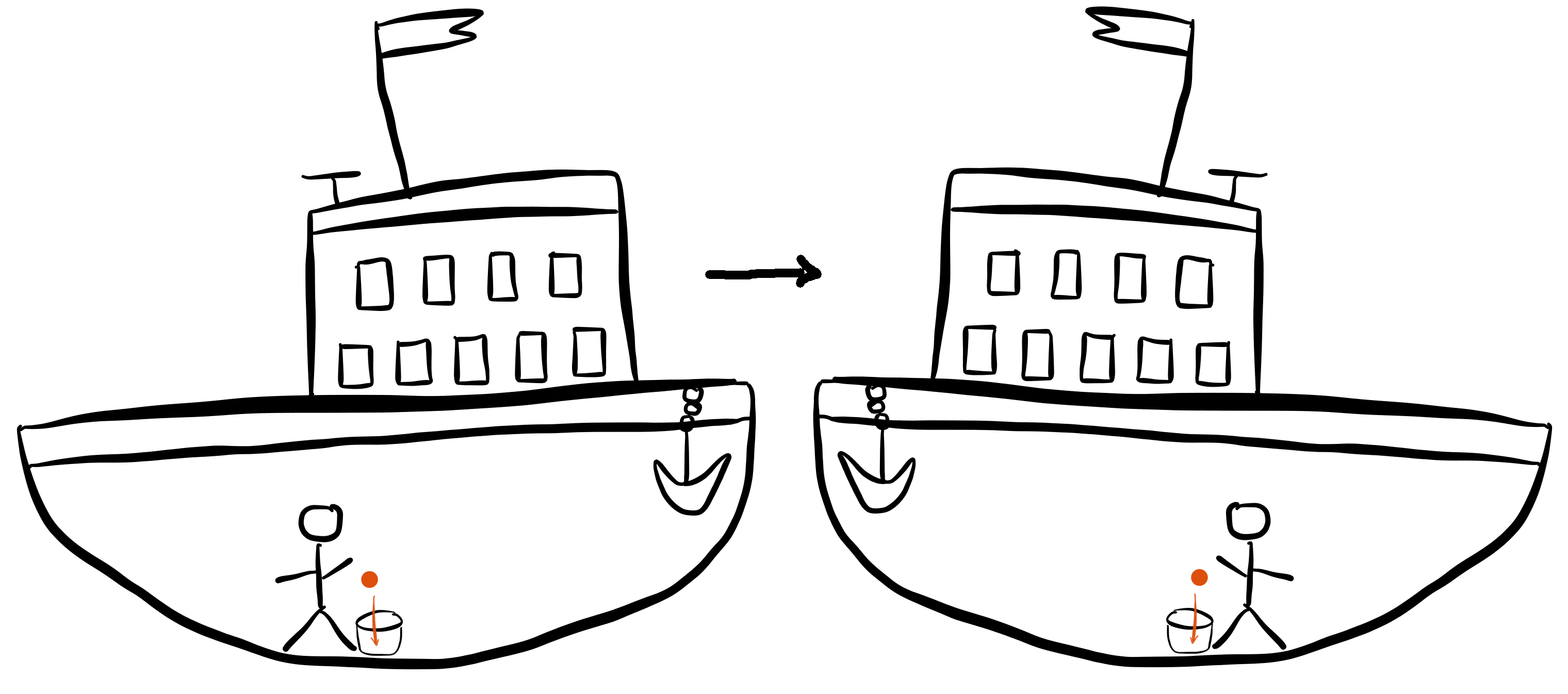}
\end{center}

Therefore, the subsystem has a global rotational symmetry. The crucial point is that, in principle, it would be possible to detect a difference, for example by bringing the subsystem in contact with the outside world. In this sense, we say that the ship and the rotated ship are two distinct states.

In contrast, there is no way we could ever detect a global rotation of the whole universe. This is why it is essential to talk about subsystems in this context \cite{Wallace2014-GREECO}. Moreover, in physics we always consider sufficiently isolated subsystems, even if this is not explicitly stated and mathematically we take the limit $|x|\to \infty$.

Now what about local transformations? 

Our physicist inside the ship would immediately notice if we rotate only a part of the ship, e.g., a specific apparatus.

\begin{center}
    \includegraphics[width=0.9\textwidth]{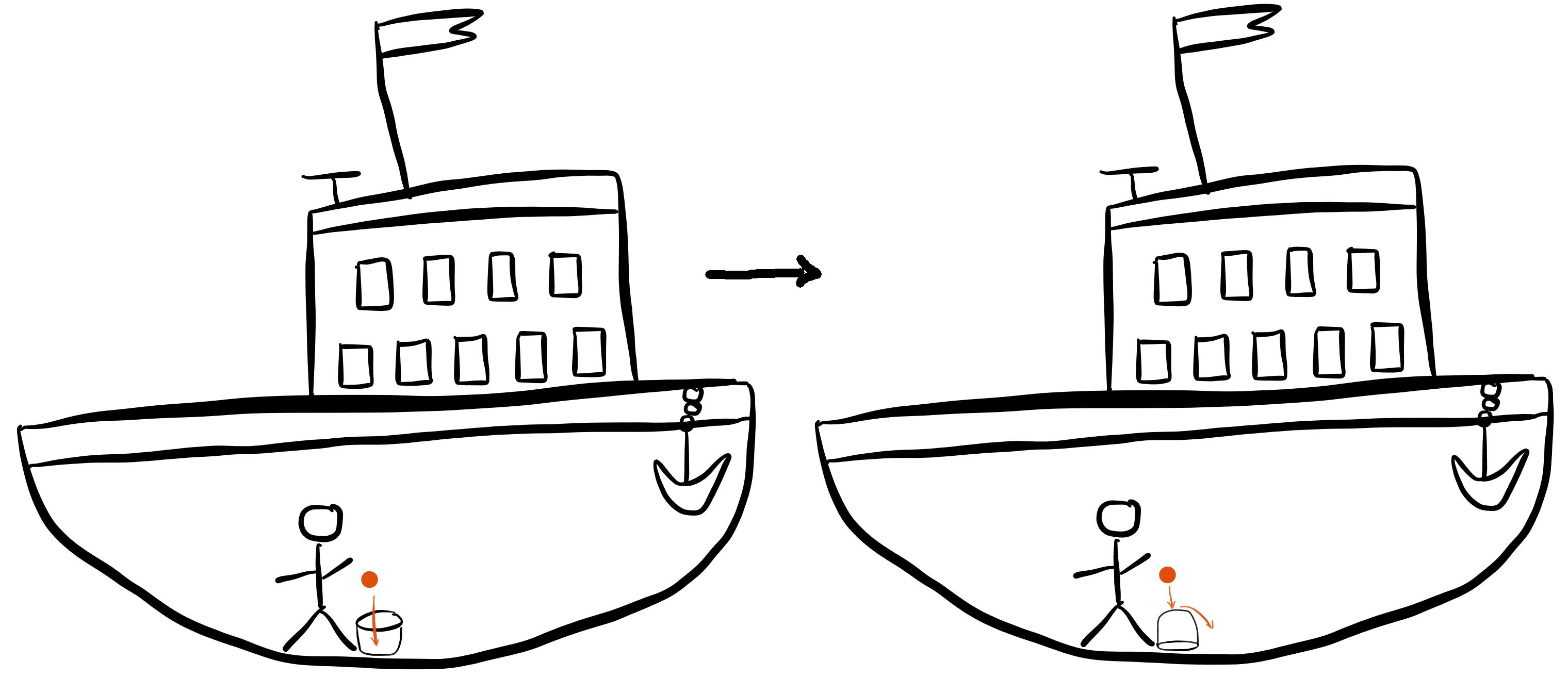}
\end{center}

Therefore, we don't have local rotational symmetry here. Similar thought experiments will be important later in the context of gauge theories where often special emphasis is put on local transformations. Already at this point we can note that while global symmetries are common, local symmetries are rarely observed in nature.\footnote{An example of a system with a local symmetry is a grid of completely black balls. Here, we can rotate each ball individually and it would be impossible to tell the difference.}

\subsection{Active vs. Passive Transformations and Symmetries vs. Redundancies}
\label{sec:activepassive}

So far, we only considered active transformations. For example, we discussed what happens when we rotate Galileo's ship. These are real physical transformations.

However, there is also a different kind of transformations, called passive transformations. A passive transformation is a change in how we \textit{describe} a given system.

\begin{center}
    \includegraphics[width=0.7\textwidth]{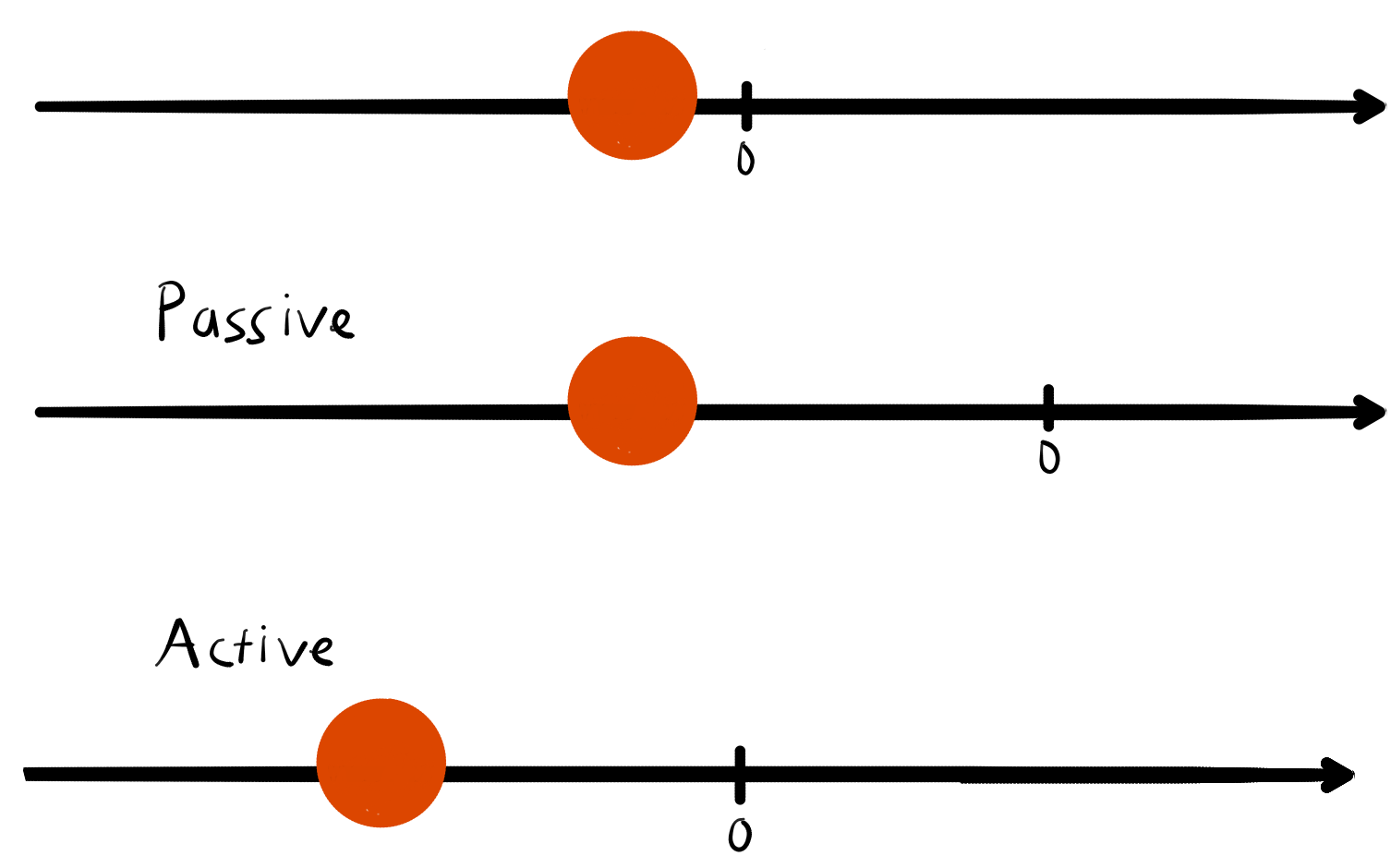}
\end{center}

For example, we can describe Galileo's ship using curvlinear coordinates or a rotated coordinate system. Such a change in how we describe a system has, of course, never any physical effect.\footnote{If we write down the equations describing our system appropriately this becomes immediately clear. We will discuss this explicitly in Section~\ref{sec:gaugetheorymathematically}.} 

Passive transformations relate different \textit{descriptions} of the same physical situation, while active transformations relate different physical situations. A lot of confusion surrounding gauge symmetries can be traced back to confusion about these two kinds of transformations.\footnote{Maybe it would be helpful to use the notions real transformation and coordinate transformation instead of active transformation and passive transformation. }

Since it is important to distinguish between passive and active transformations, we also need to distinguish between invariance under active transformations and invariance under passive transformations. Concretely, we call invariance under passive transformations a \textbf{redundancy} and invariance under active transformations a \textbf{symmetry}. This distinction is essential because a symmetry is a real feature of a system while a redundancy is only a feature of our description.

As we will discuss in detail below, we can make the description of any system invariant under any transformation by introducing appropriate mathematical bookkeepers. But this, of course, does not mean that we can make the system more symmetric this way. Instead, this only demonstrates that we can introduce redundancies by expanding our formalism. 

We can distinguish active from passive transformations by using the behavior of the bookkeepers. While the bookkeepers adjust automatically whenever we perform a passive transformation (since the system has to remain invariant), for active transformations this is not necessarily the case (since active transformations can lead to physically different states). This is a key observation which allows us to decide which gauge transformations are to be understood in an active sense and which in a passive sense.\footnote{From a purely mathematical point of view we can imagine that the bookkeepers change automatically whenever we perform an active transformation, too. Therefore, this is something which needs to be determined experimentally. }

A key insight is that the bookeepers can be more than purely mathematical tools which make sure that our description is invariant under passive transformations. They can also provide a background structure which influences the dynamics within the system. Moreover, they can change dynamically and in this sense, become active actors themselves. If this is the case, we are dealing with a gauge theory. However, irrespective of their role within the system, the remarks made above about their behavior under active and passive transformations remain valid.

Now, after these preliminary remarks we can start to discuss gauge symmetries. In the following section, we introduce all relevant notions in the context of a simple toy model.

\section{Gauge Symmetry}

\label{sec:gaugesymintuitively}

The toy model we will use in the following describes a simplified financial market. It consist of several countries and the basic process we try to describe is that money and goods can be traded and carried around. This setup is arguably the simplest setup where a gauge symmetry shows up. The connection between financial markets and gauge symmetries was first put forward in Ref.~\cite{Ilinski:1997tj} and later popularized in Ref.~\cite{doi:10.1119/1.19139,Maldacena:2014uaa}.

First of all, let's imagine that we have a common currency in several countries. For concreteness, we call this currency Euro and the countries Germany, France, the United Kingdom and Italy.

In addition, we consider this subsystem of the whole world isolated and assume that there is a trader who only does business within these countries. 
A crucial observation is now that this subsystem has a \textbf{global gauge symmetry} since the absolute value of fiat money is, in general, not determined.\footnote{The notion global gauge symmetry is defined more precisely in Section~\ref{sec:gaugesymmathematically}.} We can shift the currency or alternatively, all prices without any physical effect.\footnote{At least in our toy model changing the value of the currency has no effect, since we imagine that as a result simply all prices and wages are automatically adjusted. Of course, in the real world there could be psychological effects since people get used to certain prices.} 

\begin{center}
    \includegraphics[width=0.7\textwidth]{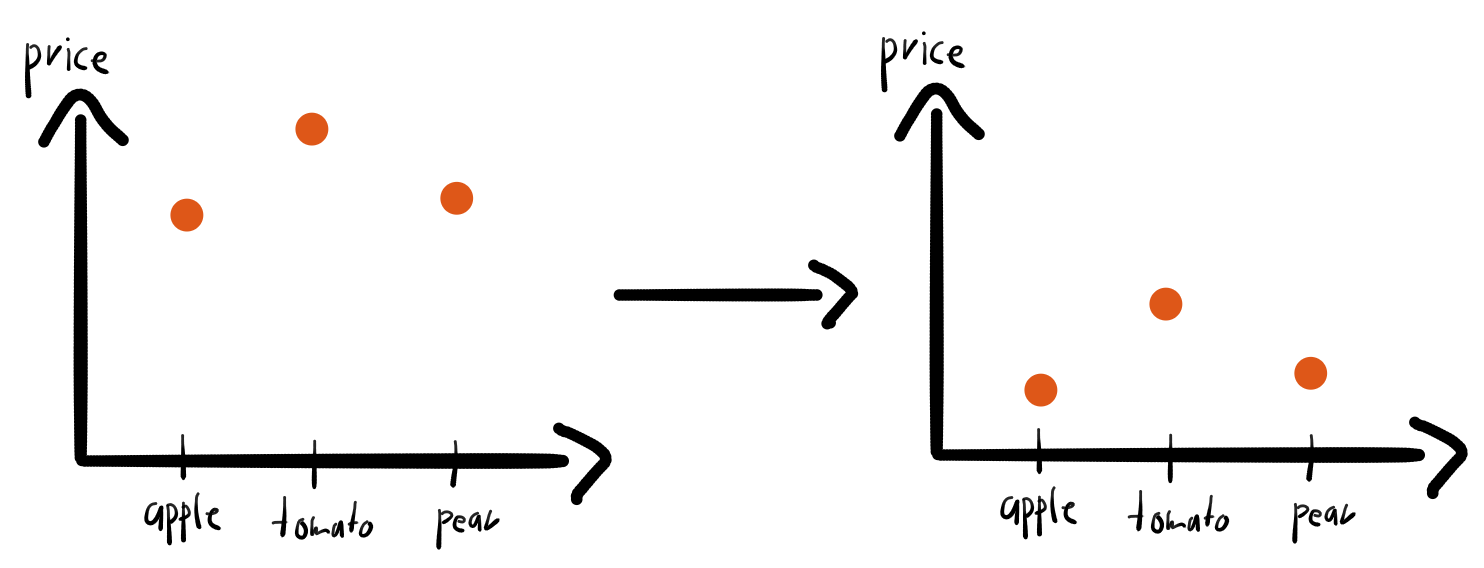}
\end{center}

To understand this, imagine that a trader sells three tomatoes at 1€ each and then uses this money to buy six apples at 0.5€ each. 

\begin{center}
    \includegraphics[width=0.5\textwidth]{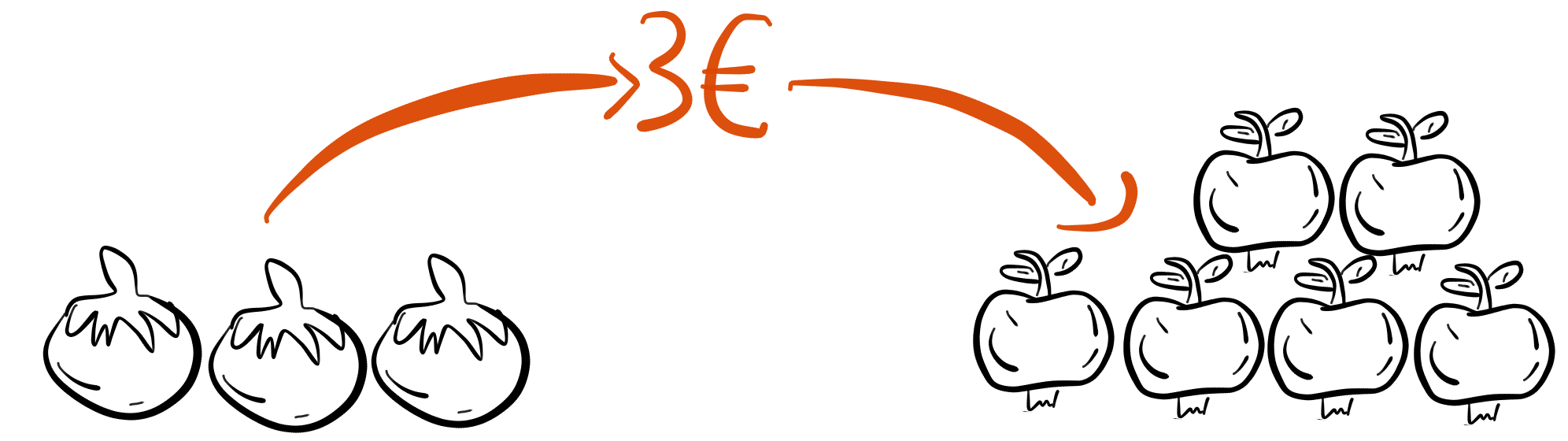}
\end{center}

Now the end result of such a process is completely unchanged if the government decides to adjust the prices of everything such that the value of each Euro drops by a factor of ten. Afterwards, the trader only gets 0.1€ for each tomato but is then able to buy apples at 0.05€. So again, our trader starts with three tomatoes and ends up with six apples.

\begin{center}
    \includegraphics[width=0.5\textwidth]{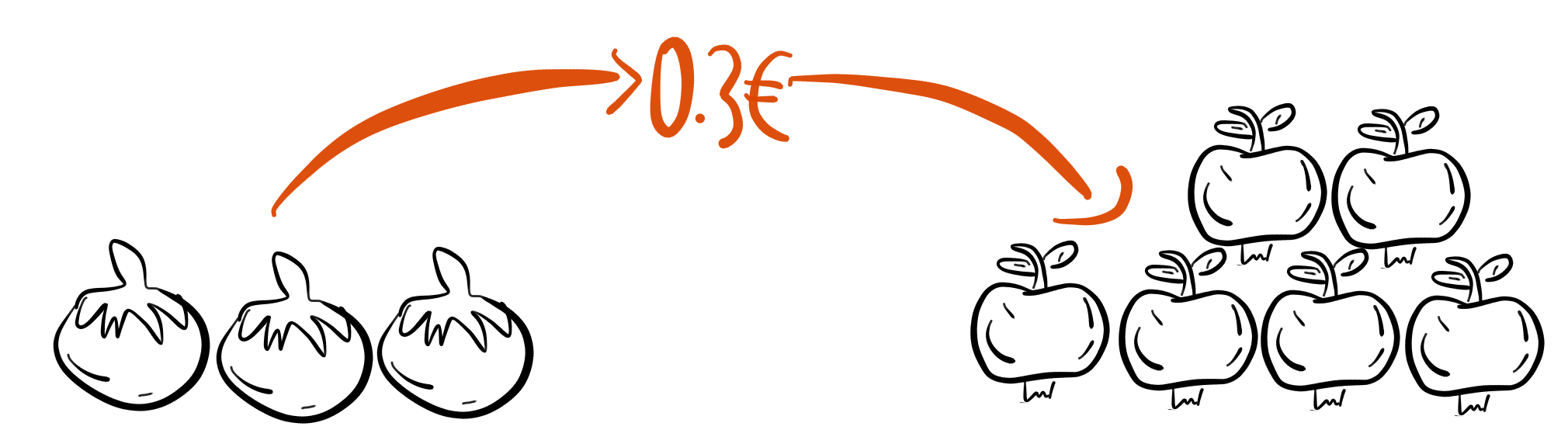}
\end{center}

In the real world we are dealing with different currencies and we can imagine that it should be possible to adjust these local currencies freely.

\subsection{Local Gauge Symmetry}

For concreteness, we now introduce an independent local currencies in Germany, which we call Deutsche Mark (DM). Moreover, we introduce Francs (F) in France, in England Pounds (P) and Lira (L) in Italy.

This is only possible if we introduce bookkeepers which keep track of the values of the local currencies and are able to exchange one currency for another. We can then imagine that the bookkeepers always adjust their exchange rates perfectly whenever the value of a local currency changes. If this the case such changes have no noticeable effect. 

For example, let's assume that the exchange rates are
\begin{align}
 DM/P &= 1 \notag \\
 P/F &= 2 \notag \\
  F/L &=  10 \notag \\
 DM/L &= 20 \, .
\end{align}
If a trader starts with $1DM$, he can trade it for $1P$, then use it to trade it for $2F$, then trade these for $20L$ and finally trade these back for $1DM$. 

\begin{center}
    \includegraphics[width=0.4\textwidth]{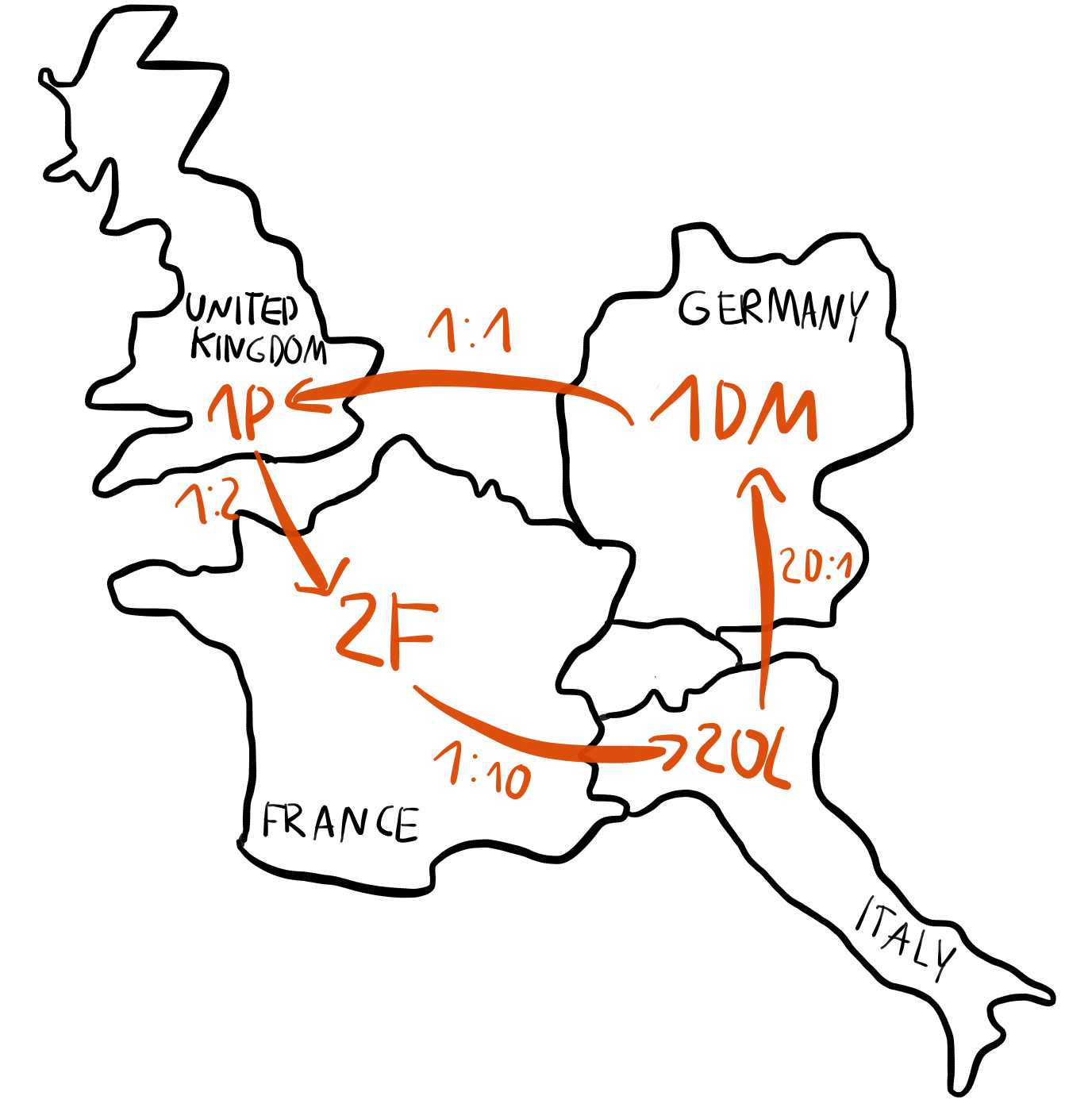}
\end{center}

Now, when the Italian government decides to drop a zero from their currency, the exchange rates become
\begin{align}
 DM/P &= 1 \notag \\
 P/F &= 2 \notag \\
 F/L &=  10 \quad   \to \quad  F/\tilde L  = F/(L/10) = 1 \notag \\
DM/L &= 20 \quad   \to \quad  DM/\tilde L = DM/(L/10) = 2 \, .
\end{align}
Therefore, if a trader starts again with $1DM$, he can trade it for $1P$, then use it to trade it for $2F$, then trade these for $2L$ and finally trade these back for $1DM$.

\begin{center}
 \includegraphics[width=0.4\textwidth]{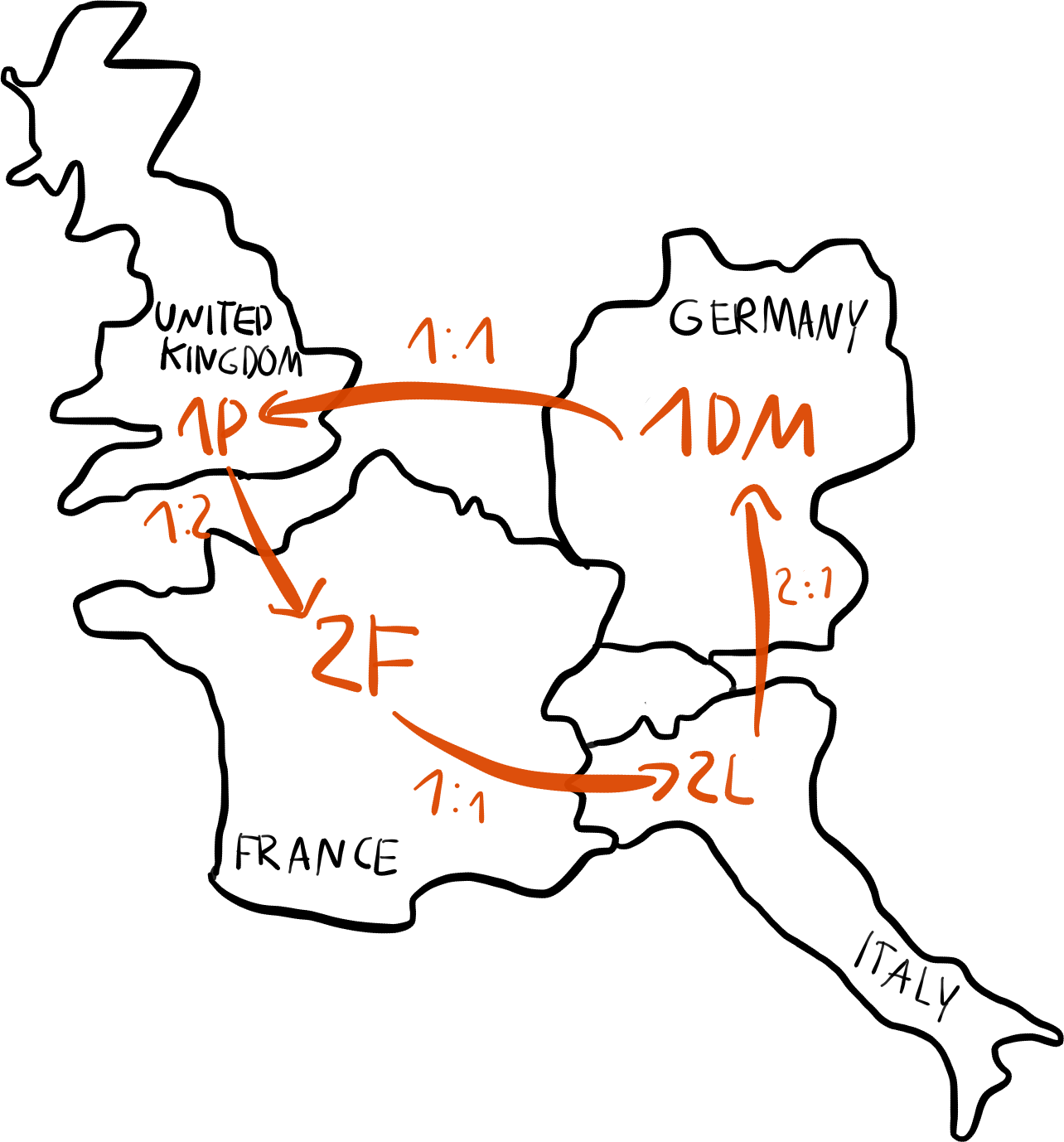}
\end{center}

A local rescaling of a given currency therefore makes no difference as long as the exchange rates get adjusted accordingly.

Using the language introduced above, we can therefore say that as soon as we introduce bookkeepers, our system is invariant under local transformations. It is conventional to call this invariance \textbf{local gauge symmetry}. However, at this point it isn't clear if we are dealing with a local symmetry and/or a local redundancy. This is something which needs to be determined experimentally. 

For concreteness, let's say that changing prices is an active transformation while shifting a currency is a passive transformation. Whether our system is invariant under active and passive transformations then depends crucially on the behavior of the bookkeepers. Since a shift of a local currency is merely a change of the local money coordinate system, it shouldn't make any difference. Therefore, it seems plausible that our bookkeepers adjust automatically whenever such a passive transformation happens. But for active price adjustments the situation is not so clear. If a country decides to lower its prices, this is something which clearly can change the dynamics of the financial market. Therefore, we may imagine that the exchange rates remain unaffected by active price adjustments. If this is the case, we are dealing with a local redundancy but not with a local symmetry. But, of course, we can also imagine that exchange rates get affected by price adjustments, too. Then we would be dealing with a local symmetry. 

In Section~\ref{sec:gaugesymphysics}, we will revisit this issue by discussing concrete experiments.

But first, we need to talk about why we care about gauge symmetry at all.

\section{Gauge Theory Intuitively}
\label{sec:gaugetheoryintuitively}

So far, our bookkeepers are purely mathematical ingredients which we introduced to make our description invariant under local transformations. In our finance toy example, it's easy to imagine that bookkeepers can influence the dynamics of a system and even become dynamical actors on their own. 

First of all, we can imagine that there are imperfections in the exchange rates. If this is the case, we can imagine that a trader starts to trade currencies since this can be a lucrative business. And this is an explicit example of how our bookkeepers can influence the dynamics of the system.

For example, let's imagine the exchange rates are as follows
\begin{align}
DM/P &= 1 \notag \\
 P/F &= 2 \notag \\
  F/L &=  10 \notag \\
 DM/L &= 10 \, .
\end{align}
Now a trader is able to earn money simply by trading currencies. If he starts with $1DM$, he can trade it for $1P$, then use the pound to trade it for $2F$, then trade these for $20L$ and finally trade these for $2DM$. 

\begin{center}
    \includegraphics[width=0.4\textwidth]{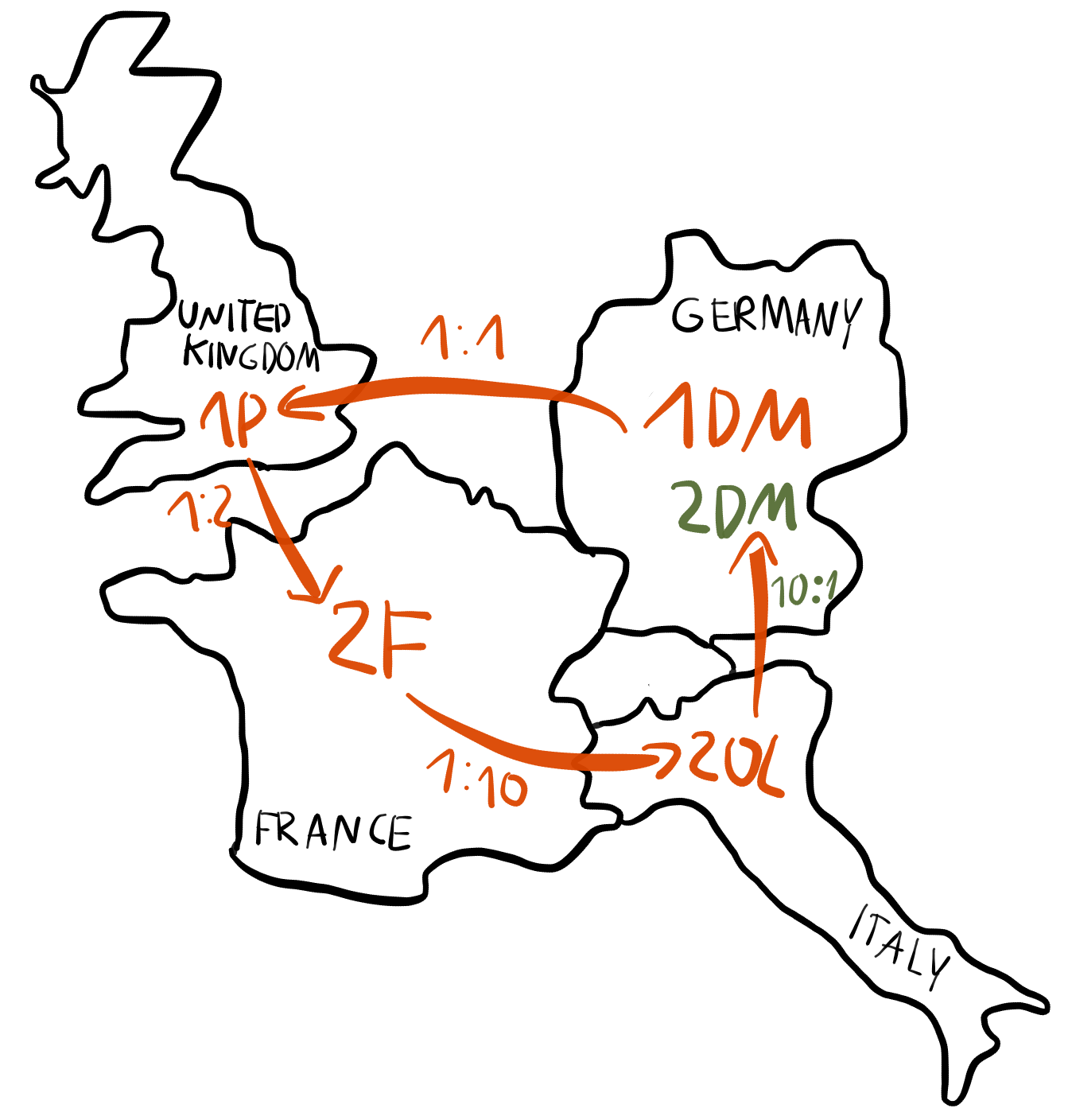}
\end{center}

In the financial world this is known as an \textbf{arbitrage opportunity}.\footnote{Arbitrage is a risk-free opportunity to earn money. In geometrical terms, we can understand the corresponding mathematical quantity as the curvature of our internal money space. } 

But wait, the exchange rates depend on the values of the local currencies which we can change at will $\ldots $ does this mean that the amount of money our trader earns depend on the these arbitrary choices?

First of all, as mentioned above, our exchange rates indeed do change when, for example, Italy decides to drop a zero from their currency. We need to take such a change of the local money coordinate system $L \to \tilde L = L /10$ consistently into account and this means we need to adjust the exchange rates accordingly:\footnote{As mentioned above, we treat dilations of a currency as passive transformations. Therefore, they are not allowed to have any effect on the system and this requires that our bookkeepers adjust accordingly.}
\begin{align}
 DM/P &= 1 \notag \\
 P/F &= 2 \notag \\
 F/L &=  10 \quad   \to \quad  F/\tilde L  = F/(L/10) = 1 \notag \\
DM/L &= 10 \quad   \to \quad  DM/\tilde L = DM/(L/10) = 1 \, .
\end{align}
However, the amount of money our trader earns is unchanged by such a re-scaling! 

If he starts again with $1DM$, he can still trade it for $1P$, then for $2F$, then trade these for $2L$ and finally trade these for $2DM$. The final result is the same as before.

\begin{center}
    \includegraphics[width=0.4\textwidth]{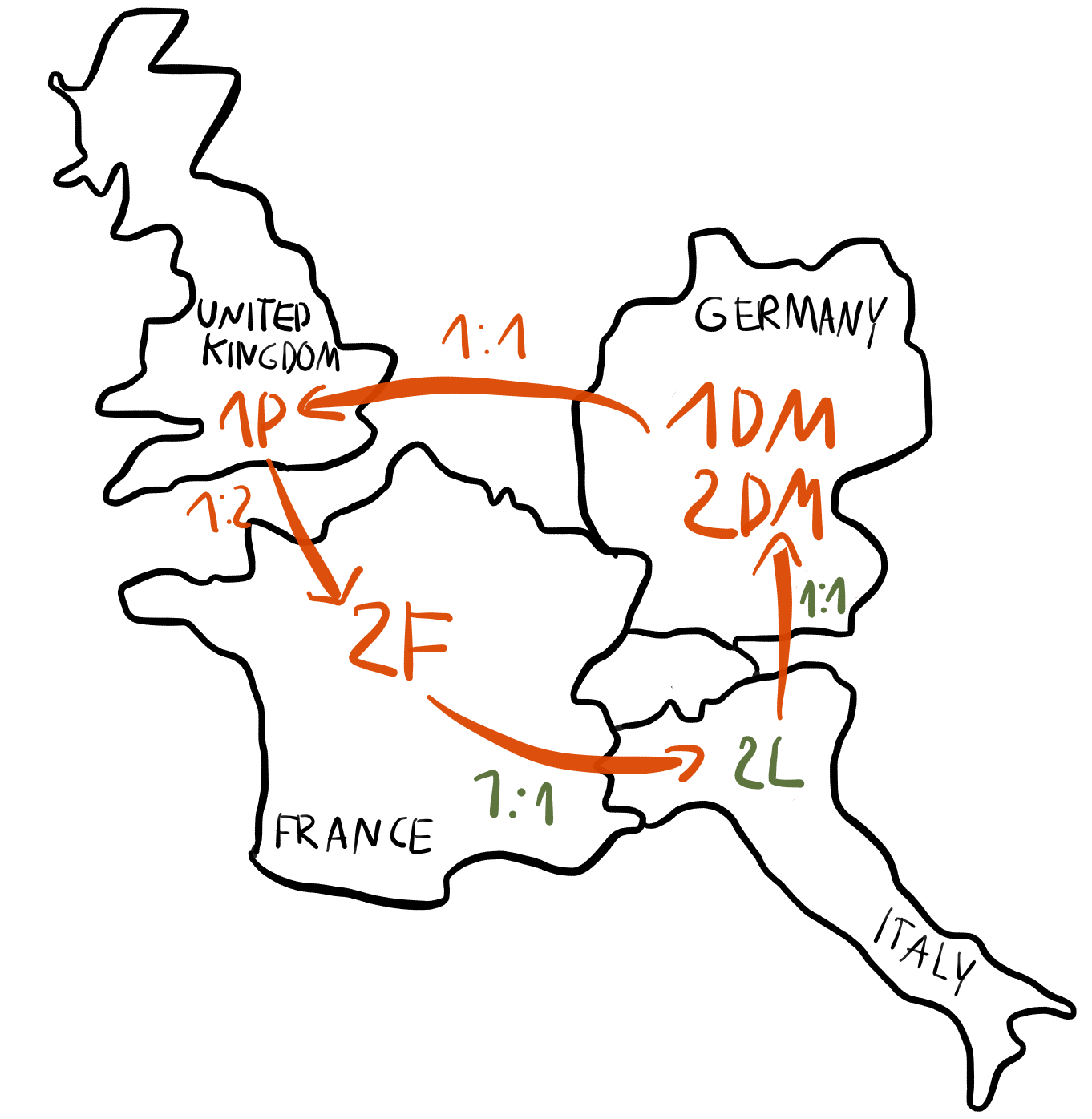}
\end{center}

What we have learned here is a crucial aspect of every \textbf{gauge theory}. An important task is to find quantities which do not depend on local conventions. For example, the opportunity to earn risk-free money is independent of how we choose our local coordinate system. 

The crucial point in these examples is that the situation in our toy model is a very different one whether there is an arbitrage opportunity or not. When there is an arbitrage opportunity, our bookkeepers are more than mere mathematical tools since their exchange rates actively shape the dynamics within the system.\footnote{In physics the situation with an arbitrage opportunity corresponds to a system with nonzero curvature. But this curvature is not necessarily a dynamical object.}

However, so far, our bookkeepers are still not dynamical. The exchange rates don't change and while the bookkeepers have a real effect on the system they only yield the static background in front of which all dynamical actors do their business.

But we can imagine that our bookkeepers become dynamical actors, too. In the real world, banks calculate exchange rates, but they are also institutions that live in the real world, not only in our description of financial markets and make decisions dynamically.

Promoting bookkeepers to dynamical objects which follow their own rules turns our model of the financial market into a gauge theory.\footnote{In physics a dynamical object is something with its own equation of motion.} 

While we can always introduce local redundancies into our description of a given system, we are only dealing with a gauge theory when the bookkeepers are dynamical objects in the system and not only artifacts of our description. The defining feature of a gauge theory is that the bookkeepers, which are necessary to make the description of the system invariant under local passive transformations, are dynamical actors that shape the dynamics of the system. In practice this means here that exchange rates are adjusted dynamically depending on what else happens in the system.

To discuss all this in more concrete mathematical terms, we need to describe our toy model in mathematical terms.

\subsection{Mathematical Description of the Toy Model}

Mathematically, we imagine that our \textbf{countries} live on a lattice. Each point on the lattice is labelled by $d$-numbers: $\vec n = ( n_1, n_2, \cdots , n_d)$. In other words, each country can be identified by a vector $\vec n$ which points to its location.\footnote{The notation in this section is adapted from Ref~\cite{Maldacena:2014uaa}. } 

We can move from one country to a neighboring country by using a \textbf{basis vector} $\vec{e}_i$, where $i$ denotes the direction we are moving. For example, $\vec {e}_2=( 0,1,0\cdots  , 0 )$.

We denote the \textbf{exchange rates} between the country labelled by the vector $\vec{n}$ and its neighbor in the $i$-direction by $R_{\vec n , i }$. For example, if the country at the location labelled by $\vec n$ uses Deutsche Marks and its neighbor in the $2$-direction uses Francs, $R_{\vec n , 2 }$ tells us how many Francs we get for each Deutsche Mark.

\begin{center}
 \includegraphics[width=0.4\textwidth]{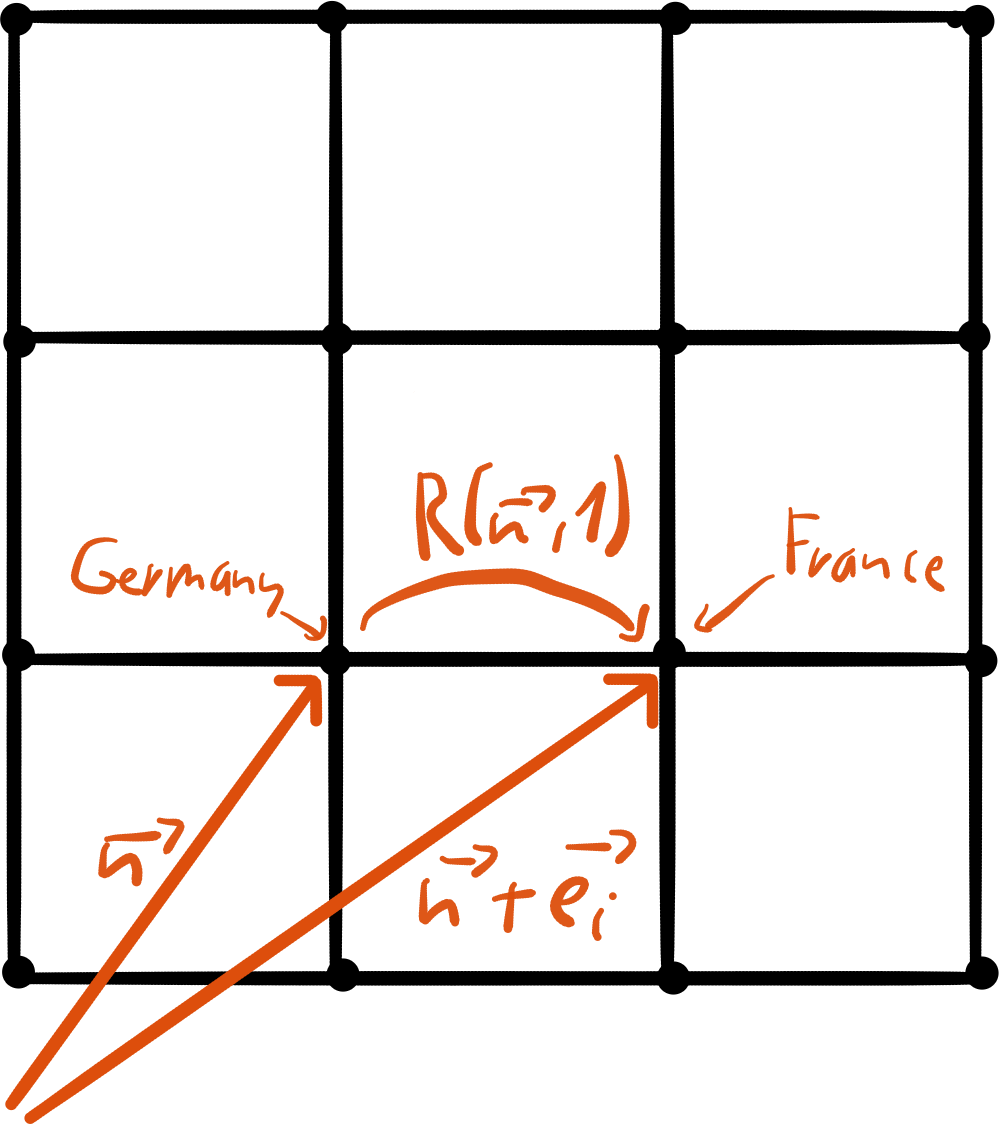}
\end{center}

In physics, we usually introduce the corresponding logarithm, here of the exchange rates
\begin{equation}
     R_{\vec n , i }  \equiv  e^{ A_i ( \vec n ) } \, ,
\end{equation}
where $A_i ( \vec n ) \equiv \ln(R_{\vec n , i }) $.

The next ingredient that we need is a notation for gauge transformations. In our toy model a gauge transformation is a change of currencies and directly impacts the exchange rates. We use the notation $f(\vec n)$ to denote a change of the currency in the country at $\vec n$ by a factor $f$. In addition, we again introduce the corresponding logarithm
\begin{equation}
     f(\vec n) \equiv e^ { \epsilon(\vec n) }\, .
\end{equation}
In general, when we perform such a gauge transformation in the country labelled by $\vec{n}$ and also in the neighboring country in the $i$-direction, the corresponding exchange rate changes as follows\footnote{The currency in the country at $\vec{n}$ gets multiplied by $f_{\vec n}$ and the currency in its neighbor in the $i$-direction by $f_{\vec n + \vec e_i}$. Therefore, the exchange rate gets modified by the ration of these two factors. }
\begin{equation}
     R_{\vec n , i } \to  {   f_{\vec n + \vec e_i }  \over f_{\vec n}}   R_{\vec n , i } \, .
\end{equation}
In terms of the logarithms this equation reads
\begin{align}
     R_{\vec n , i }= e^{ A_i ( \vec n ) } \to  &{   f_{\vec n + \vec e_i }  \over f_{\vec n}}   R_{\vec n , i } \notag \\
     &= {   e^ { \epsilon(\vec n + \vec e_i ) }  \over e^{\epsilon(\vec n)}}   e^{ A_i ( \vec n ) }  \notag \\
     &=   e^{ A_i ( \vec n ) + \epsilon(\vec n + \vec e_i ) -  \epsilon(\vec n) } 
\end{align}
and we can conclude
\begin{align} \label{eq:gaugetrafofield}
    A_i(\vec n) \to A_i(\vec n) + \epsilon(\vec n + \vec e_i) - \epsilon(\vec n) \, .
\end{align}
We learned above that an import aspect of the system is whether arbitrage possibilities exist. An arbitrage opportunity exists when we can trade currencies in such a way that end up with more money than we started with. But we can only make such a statement when the starting currency and the final currency are the same. Only then we can be certain whether the final amount of money is larger than the initial amount. Therefore, we need to trade money in a loop. 

The total gain we can earn by following a specific loop can be quantified by
\begin{equation}
    G =   R_{\vec n , i} R_{\vec n + \vec e_i , j } { 1 \over R_{\vec n +  \vec e_j  , i } }{ 1 \over 
R_{ \vec n  , j } } \, .
\end{equation}
When this \textbf{gain factor} is larger than one, we can earn money by trading money following the loop, if it is smaller than one we loose money.

To understand the definition of the gain factor imagine that we start with $1DM$. We trade it for Pounds and $R_{\vec n , 1}=1$ tells us that we get $1P$. Afterwards, we trade our Pounds for Francs and 
$R_{\vec n + \vec e_1 , 2 }=2$ tells us that we get in total $2F$. Afterwards, we trade our France for Lira.  $R_{ \vec n +  \vec e_2  , 1 }=10$ tells us that we get $1/10$ Franc for each Lira. Hence, we have to calculate $2F / R_{ \vec n +  \vec e_2  , 1 }=20L$. Finally, we use that $R_{ \vec n  , 2 } =10$ tells us that we get $10L$ for each Deutsche Mark and therefore calculate $20L/R_{ \vec n  , 2 }=2DM$.\footnote{The crucial point here is that our exchange rates $R_{\vec n , i}$ always tell us how many of the currency at the neighboring country in the $i$-direction we get for each unit of the local currency at the country at $\vec{n}$. Hence, we sometimes have to divide by the corresponding exchange rate to calculate the resulting amount of a new currency. }

\begin{center}
    \includegraphics[width=0.4\textwidth]{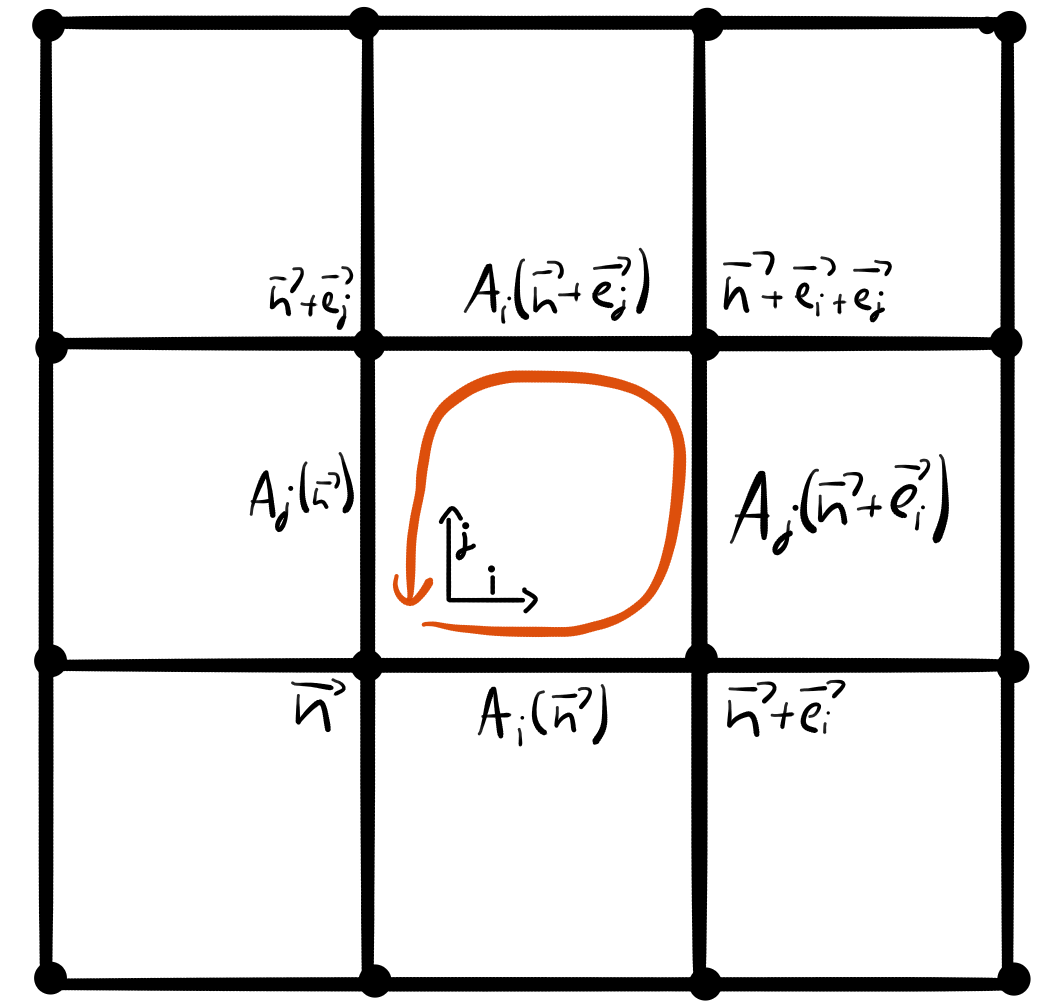}
\end{center}
Once more we introduce the corresponding logarithm
\begin{equation}
    G  \equiv e^{F_{ij}(\vec n ) }  
\end{equation}
and again, we can rewrite our equation in terms of the logarithms\footnote{In physics $F_{ij}$ is directly related to components of the \textbf{magnetic field}. For example, $F_{12}=B_3$.}
\begin{equation} \label{eq:fieldstrengthtesnor}
    F_{ij}(\vec n) = A_j(\vec n + \vec e_i) - A_j(\vec n ) -  [ A_i(\vec n + \vec e_j) - A_i(\vec n) ] \, .
\end{equation}
A crucial consistency check is that $G$ and $F_{ij}$ are unchanged by gauge transformations. We already argued above that an arbitrage opportunity is something real and thus cannot depend on local conventions. Quantities like this are usually called \textbf{gauge invariant}. So in words, $G$ and $F_{ij}(\vec n)$ encode what is physical in the structure of exchange rates.\footnote{A single exchange rate $A_i(\vec{n})$ is gauge dependent and can therefore, for example, be set to zero simply by changing local money coordinate systems.} Moreover, an important technical observation is that $F_{ij}(\vec n)$ is antisymmetric: ${F_{ij}(\vec n) = - F_{ji }(\vec n)}$, which follows directly from the definition. 

So far, we only talked about spatial exchange rates. However, there are also temporal exchange rates, i.e. interest rates. A clever trick to incorporate this is to introduce time as the zeroth-coordinate like we do it in Special Relativity. In other words, in addition to specific locations (countries) our lattice now contains copies of these locations at different points in time. This means a point of the lattice is specified by $d+1$ coordinates $\vec n = (n_0, n_1, n_2, \cdots , n_d)$ and the zeroth component indicates the point in time. 

Then, Eq.~\ref{eq:fieldstrengthtesnor} reads\footnote{The non-vanishing components of $F_{\mu \nu}(\vec n)$ with either $\mu =0$ or $\nu =0$ are directly related to what we call electric field in physics. For example, $F_{01}=E_1$.}
\begin{equation} \label{eq:fieldstrengthtesnorcovariant}
    F_{\mu \nu}(\vec n) = A_\mu(\vec n + \vec e_\mu) - A_\nu(\vec n ) -  [ A_\mu(\vec n + \vec e_\nu) - A_\mu(\vec n) ] \, ,
\end{equation}
where previously $ij \in \{1,2,\ldots,d \}$ and now $\mu, \nu \in \{0,1,2,\ldots,d \}$.

In the continuum limit, where the lattice spacing goes to zero, Eq.~\ref{eq:gaugetrafofield} becomes\footnote{To understand this take note that in Eq.~\ref{eq:fieldstrengthtesnorcovariant} we get in this limit the difference quotient.}\begin{align} \label{eq:gaugetrafofieldcovariant}
    { A}_\mu ( x_\mu) \to { A}_\mu (x_\mu) + { \partial \epsilon \over \partial x^\mu } 
\end{align}
and Eq.~\ref{eq:fieldstrengthtesnorcovariant} reads
\begin{equation} \label{eq:fieldstrengthtesnorcovariantconti}
    { F}_{\mu \nu}(x_\mu) \equiv  { \partial { A}_\nu \over \partial x^\mu } -  { \partial { A}_\mu \over \partial x^\nu  } \, .
\end{equation}

We can not only earn money by trading money itself, but also by trading goods like, for example, copper. Depending on the local prices it can be lucrative to buy copper in one country, bring it to another country, sell it there, and then go back to the original country to compare the final amount of money with the amount of money we started with.

The gain factor for such a process is given by 
\begin{equation} \label{eq:goodgain}
      g =  { p(\vec n + \vec e_i) \over p(\vec n) R_{\vec n, i }  } \, .
\end{equation}
To understand this definition, imagine we start with $10DM$ and the price for one kilogram of copper in Germany is $p(\vec n )=10DM$. This means we can buy exactly $1$ kilogram of copper. Then we can go to the neighboring country and sell our copper for, say, $30F$ since $p(\vec n + \vec e_1)=30F$. Afterward, we can go back to Germany and exchange our $30F$ for $15DM$ since, say, $R_{\vec n, i }=0.5$. Therefore, we have made in total $5DM$ here.  Again, a gain factor larger than one means that we earn money and a gain factor smaller than one that we loose money.\footnote{If you are unsure which quantity goes in the numerator and which in the denominator, ask yourself: Would it increase our profit if the given quantity is larger? If the answer is yes we write in the numerator, if not in the denominator. For example, a higher price of copper in France certainly increases our profit. Hence $p(\vec n + \vec e_i)$ is written in the numerator. Similarly, a higher price of copper in Germany would lower our profit and therefore, we write in the denominator. }

\begin{center}
    \includegraphics[width=0.4\textwidth]{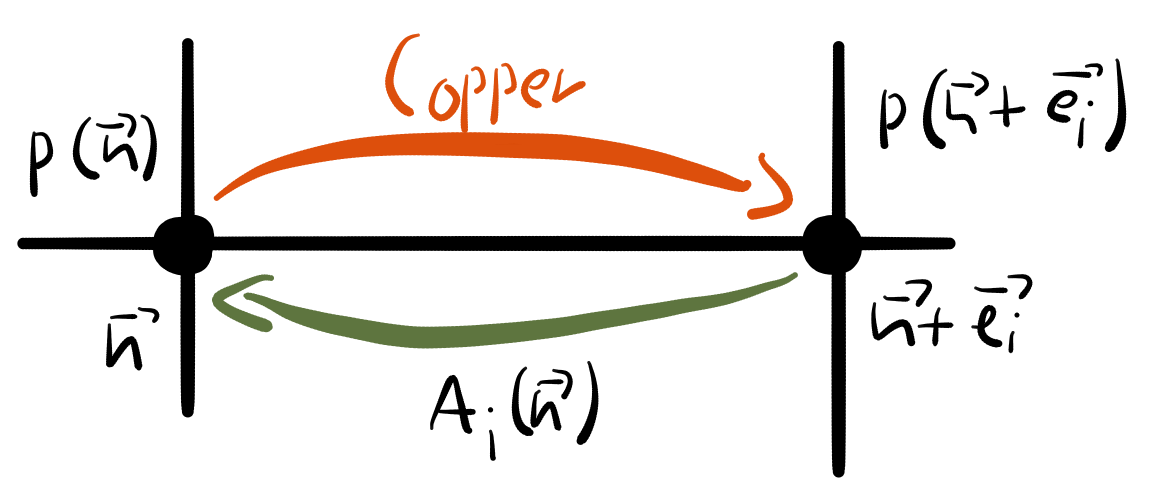}
\end{center}

Once more, we introduce the corresponding logarithm
\begin{equation}
      g \equiv  e^{ J_i(\vec {n}) }
\end{equation}
and Eq.~\eqref{eq:goodgain} then read in terms of the corresponding logarithms
\begin{align}
g &= { p(\vec n + \vec e_i) \over p(\vec n) R_{\vec n, i }  } \notag \\
\therefore  \quad e^{ J_i(\vec {n}) } &= {  e^{\varphi(\vec n+ \vec e_i)}   \over e^ {\varphi(\vec n)} e^{ A_i ( \vec n ) }  }  \notag \\
\therefore  \quad  J_i(\vec{n}) &= \varphi(\vec n  + \vec e_i) - \varphi( \vec n ) - A_i(\vec n) \, .
\end{align}
The amount of money we earn depends on the amount of copper we carry around. Thus, in general, we have
\begin{align}
 J_i(\vec{n}) &= q \Big( \varphi(\vec n  + \vec e_i) - \varphi( \vec n ) - A_i(\vec n)\Big) \, ,
\end{align}
where $q$ is related to the amount of copper involved in the trade.
Completely analogous to what we did above, we can generalize this formula for situations that involve time by replacing $i \in \{1,2,\ldots,d\}$ with $\mu \in \{0,1,2,\ldots,d\}$).
\begin{align} \label{eq:conservedcurrenttoymodel}
   J_\mu(\vec{n}) &= q \Big( \varphi(\vec n  + \vec e_\mu) - \varphi( \vec n ) - A_\mu(\vec n) \Big) \, .
\end{align}
In the continuum limit, Eq.~\ref{eq:conservedcurrenttoymodel} becomes
\begin{align} 
J_\mu(\vec{n}) &= q \Big( \frac{\partial \varphi}{\partial x_\mu} - A_\mu(\vec n) \Big ) \, .
\end{align}
In addition to the interpretation as a gain factor, there is another way how we can look at the four quantities $J_\mu(\vec{n})$.

As mentioned above, the amount of money we can earn in a copper trade $J_\mu$ is proportional to the amount of copper involved. Hence, we can use $J_\mu$ as a measure, for example, of the amount of copper that flows between countries. In the trade described by $J_i(\vec{n})$ copper is transported from the country at $\vec{n}$ to the neighboring country at $\vec{n}+\vec{e}_i$. Hence, $J_i(\vec{n})$ is a measure of the amount of copper that flows between the two countries.

The trade related to the gain factor $J_0(\vec{n})$ does not involve the exchange of copper between neighboring countries. Instead, $J_0(\vec{n})$ tells us how much money we can earn by buying copper and selling it at a later point in time in the \textit{same} country. Again, $J_0(\vec{n})$ is directly proportional to the amount of copper involved. Hence, $J_0(\vec{n})$ is a measure of the amount of copper in the country at $\vec{n}$.

An important idea in every gauge theory is that the equation of motions should not depend on local conventions. Using what we learned above, this means that the equation should only depend on $J_\mu$ and $F_{\mu \nu}$. If we then assume that the amount of copper is conserved within the system ($\partial_\mu J_\mu=0$), we can derive the inhomogeneous Maxwell equation
\begin{equation}
\partial_\nu F_{\mu \nu} = \mu_0 J_\mu \, ,
\end{equation}
where $\mu_0$  is a constant which encodes how strongly the pattern of arbitrage possibilities react to the presence and flow of copper. This equation only contains the gauge invariant quantities $J_\mu$ and $F_{\mu \nu}$, has exactly one free index ($\mu$) on both sides and we get zero if we calculate the derivative
\begin{align}
\partial_\mu J_\mu &=0 && \text{\footnotesize{(conservation of copper)}} \notag \\
\partial_\mu \partial_\nu F_{\mu \nu} &=0 && \text{\footnotesize{(\(F_{\mu \nu}\) is antisymmetric)}} \notag \, .
\end{align}

Now that we've introduced a mathematical notation to describe the finance toy model, we can move on and discuss the simplest instances of gauge symmetries in physics.

\section{Gauge Symmetry in Physics}
\label{sec:gaugesymphysics}

\subsection{Gauge Symmetry in Quantum Mechanics}

In our money toy model, we use the local prices $p (\vec n)$ to describe traders carrying goods like copper around. Analogously, in Quantum Mechanics we use the wave function $\Psi(x)$ to describe particles carrying electric charge like, for example, electrons. A wave function is a complex function which can be written in polar form
\begin{equation}
     \Psi(x) = R(x) \mathrm{e}^{i \varphi(x)} \, .
\end{equation}
Observables are related to products of the form $\psi^\star_i(x) \hat O \Psi(x)$, where $\hat O$ denotes, as usual, an operator. Therefore, we can multiply the wave function by a global phase factor without changing anything\footnote{This is a global transformations, analogous to, for example, a rotation of the whole subsystem. Under such a global rotation all vectors have to be rotated accordingly. Analogously, here all wave functions must be transformed, i.e. $\psi_i \to \mathrm{e}^{i\epsilon} \psi_i $, too. Moreover, take note that for a particle carrying charge $q$ we have $\Psi(x) \to \mathrm{e}^{iq \epsilon}\Psi(x) $. }
\begin{equation} \label{eq:globalphaseshift}
    \Psi(x) \to \mathrm{e}^{i\epsilon} \Psi(x) \, 
\end{equation}
since
\begin{equation} 
  \psi^\star_i(x) \hat O \Psi(x) \to  \psi^\star_i(x) \mathrm{e}^{-i\epsilon}  \hat O \mathrm{e}^{i\epsilon}  \Psi(x) = \psi^\star_i(x) \hat O \Psi(x)
\end{equation}
It is important to note that this is an observable symmetry completely analogous to, for example, the rotational symmetry of Galileo's ship. To understand this, imagine that a quantum physicist sits inside the ship which represents our subsystem. This quantum physicist performs an experiment with electrons which are injected through a small slit. We can perform a global transformation as described by Eq.~\ref{eq:globalphaseshift} using a phase shifter. To get a global shift, we phase shift the electrons before we insert them into the ship. The crucial point is now that it's impossible for the quantum physicist inside the box to find out whether we performed such a phase shift or not. Therefore, a global phase shift is indeed a symmetry. 

\begin{center}
    \includegraphics[width=1.0\textwidth]{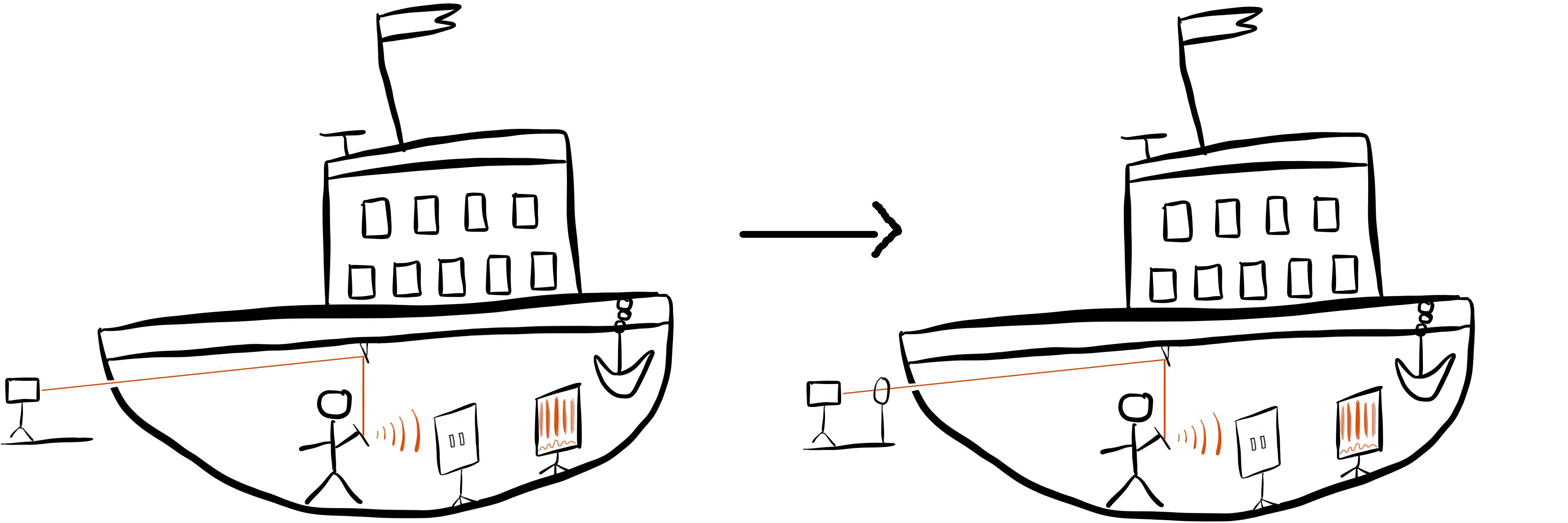}
\end{center}

Now, what about local phase shifts in Quantum Mechanics? A local phase shift is a transformation of the form \begin{equation} \label{eq:localphaseshift}
    \Psi(x) \to \mathrm{e}^{i\epsilon(x)} \Psi(x) \, ,
\end{equation}
where the transformation parameter $\epsilon(x)$ is now a function of the location $x$, i.e. no longer globally the same. Again it's important to keep in mind that this kind of transformation can be understood in an active and in a passive sense.

We can notice immediately that our description is not invariant since the momentum operator $\hat p = -i  \partial_x$ contains a derivative:\footnote{For simplicity, we restrict ourselves here to one spatial dimension and work with $\hbar =1$.}
\begin{align}
    \psi^\star_i(x) \hat p \Psi(x) \to & \psi^\star_i(x) \mathrm{e}^{-i\epsilon(x)} \hat p \mathrm{e}^{i\epsilon(x)} \Psi(x) \notag \\
    & \, = -i   \psi^\star_i(x) \mathrm{e}^{-i\epsilon(x)} \partial_x  \mathrm{e}^{i\epsilon(x)} \Psi(x) \notag \\
    & \, = -i  \psi^\star_i(x) \partial_x   \Psi(x) +  \psi^\star_i(x) \Big(\partial_x \epsilon(x) \Big)   \Psi(x) \notag \\
    &\, \neq \psi^\star_i(x) \hat p \Psi(x) \, .
\end{align}
However, if we interpret the local transformation in a passive sense, it shouldn't make any difference. Analogous to what we did in our money toy model, we can achieve this by introducing a bookkeeper $A_\mu$ which keeps track of such local changes of the phase. In particular, we replace the momentum operator, with the so-called covariant momentum operator 
\begin{equation} \label{eq:covariantmomentum}
    \hat{P} = -i  \partial_x - A_x  \, .
\end{equation}
This bookkeeper $A_x$ becomes under a local phase shift (Eq.\ref{eq:localphaseshift})\footnote{This is exactly the continuum limit of Eq.~\ref{eq:gaugetrafofield}.}
\begin{equation}
    A_x \to A_x + \partial_x \epsilon(x) \, .
\end{equation}
After the introduction of this bookkeeper our description is indeed invariant under local phase shifts.
\begin{align} \label{eq:invariance}
    \psi^\star_i(x) \hat P \Psi(x) \to & \psi^\star_i(x) \mathrm{e}^{-i\epsilon(x)} \hat{\tilde{ P}} \mathrm{e}^{i\epsilon(x)} \Psi(x) \notag \\ 
    & \, =   \psi^\star_i(x) \mathrm{e}^{-i\epsilon(x)} \Big( -i \partial_x - A_x - \partial_x \epsilon(x) \Big)  \mathrm{e}^{i\epsilon(x)} \Psi(x) \notag \\
    & \, = -i \psi^\star_i(x) \partial_x   \Psi(x) +  \psi^\star_i(x) \Big(\partial_x \epsilon(x) \Big)   \Psi(x) \notag \\
    & \quad - \psi^\star_i(x)   A_x   \Psi(x) -  \psi^\star_i(x)   \Big(\partial_x \epsilon(x) \Big)   \Psi(x)   \notag \\
    &\,  = \psi^\star_i(x) \hat P \Psi(x) \, .
\end{align}
But are local phase shifts a real symmetry of Quantum Mechanics?

To understand this, we again ask our quantum physicist inside the ship to perform an experiment with electrons. However, this time we do not perform the phase shifts globally but locally. This means, we put a phase shifter inside the ship. 

However, it's easy for the quantum physicist to find out that a phase shift did happen, even if he can't see the phase shifter directly. All he has to do is perform a double slit experiment since the result of the double slit experiment is dramatically altered by a local phase shift\cite{Aharonov:1959fk,tHooft:1980aiu}.

\begin{center}
    \includegraphics[width=1.0\textwidth]{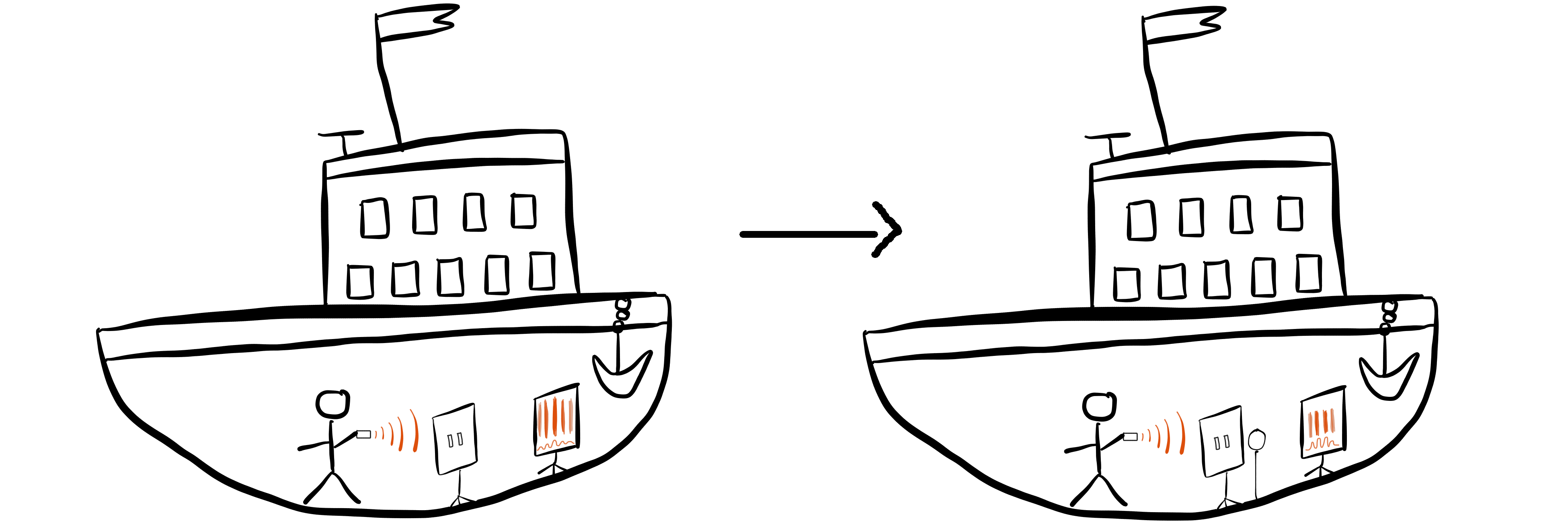}
\end{center}

Therefore, we can conclude that local phase shifts are not symmetries of Quantum Mechanics. 

Now, it's again possible that our bookkeepers $A_\mu$ become real dynamical actors, completely analogous to what we discussed for our money toy model. The theory which describes the dynamics of the bookkeepers is known as Electrodynamics.

\subsection{Gauge Symmetry in Electrodynamics}

The equations of Electrodynamics (Maxwell's equations) also posses a global symmetry
\begin{equation}
    A_\mu(x) \to A_\mu(x) + a_\mu \, ,
\end{equation}
where $a_\mu$ are arbitrary real numbers. This comes about since we can only measure potential differences.

Like in the two previous examples, the invariance under this global transformation is a real observable symmetry. To understand this, we again imagine a physicist in a ship which this time, however, is isolated from the ground. 

Charging the ship leads to a global increase of the electric potential $A_0 (x) \to A_0 (x) + \phi_0 $, while $A_i(x)$ remains unchanged. However, this change has no measurable effect inside the subsystem since $B_i =  \epsilon_{ijk} \partial_j A_k$ and $E_i = - \partial_i A_0 - \partial_t A_i$ are unchanged.

So, as long as we raise the electric potential $A_0$ globally inside the ship, it's impossible for the physicist to detect that we changed the potential.\footnote{If this is unclear, recall that, for example, a bird can walk on an uninsulated power line without any injuries. This is possible because the bird walks solely in the subsystem "power line" and the electric potential is only high relative to the potential at the ground. Alternatively, imagine Faraday sitting in a metal cage which is insulated from the ground. It's possible to electrify the cage without any notable effect inside the cage. In other words, we can raise the electric potential globally inside such a Faraday cage without inducing any measurable effect \cite{Aharonov:1959fk}. }

\begin{center}
    \includegraphics[width=1.0\textwidth]{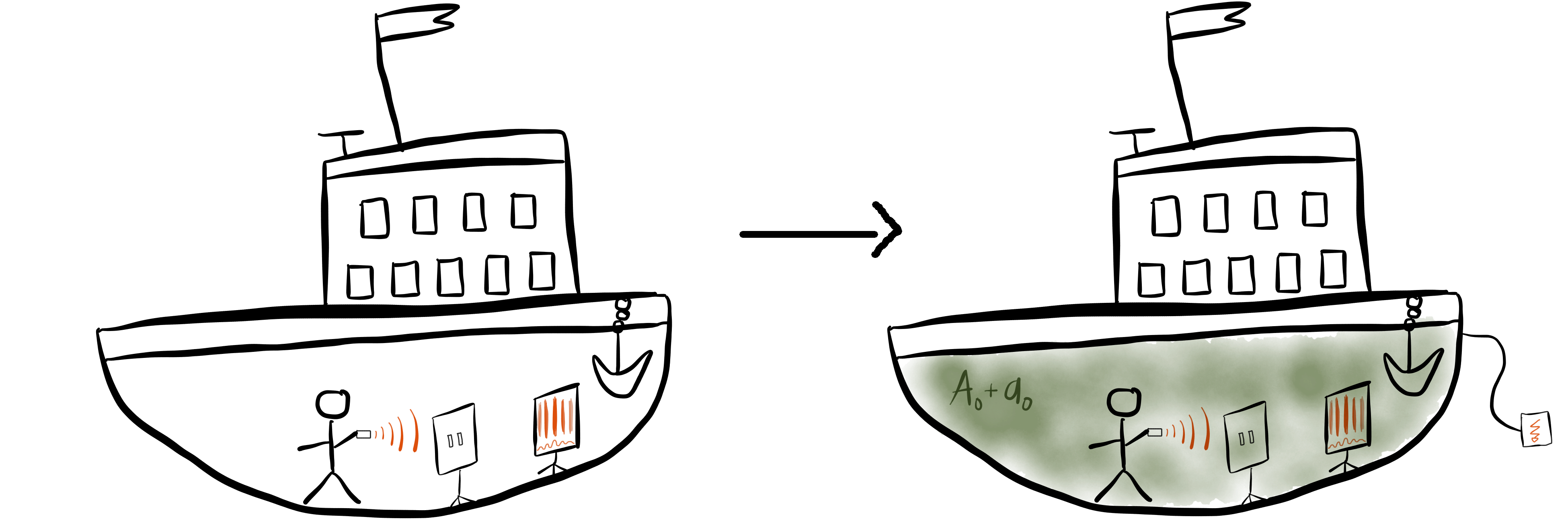}
\end{center}
Similarly to what we discussed above, we can now also argue that a local shift of the electromagnetic potential is not a symmetry of the system. To understand this, imagine that we only change the potential of a single object inside the ship. Clearly the physicist would have no problem finding this out. 

The most important point is that in Electrodynamics our bookkeepers $A_\mu$ are dynamical physical actors that can induce real physical changes. A famous example of this phenomenon is the Aharonov-Bohm effect, where a nonzero $A_\mu$ induces a phase shift in the wave function \cite{Aharonov:1959fk}. 

\begin{center}
    \includegraphics[width=1.0\textwidth]{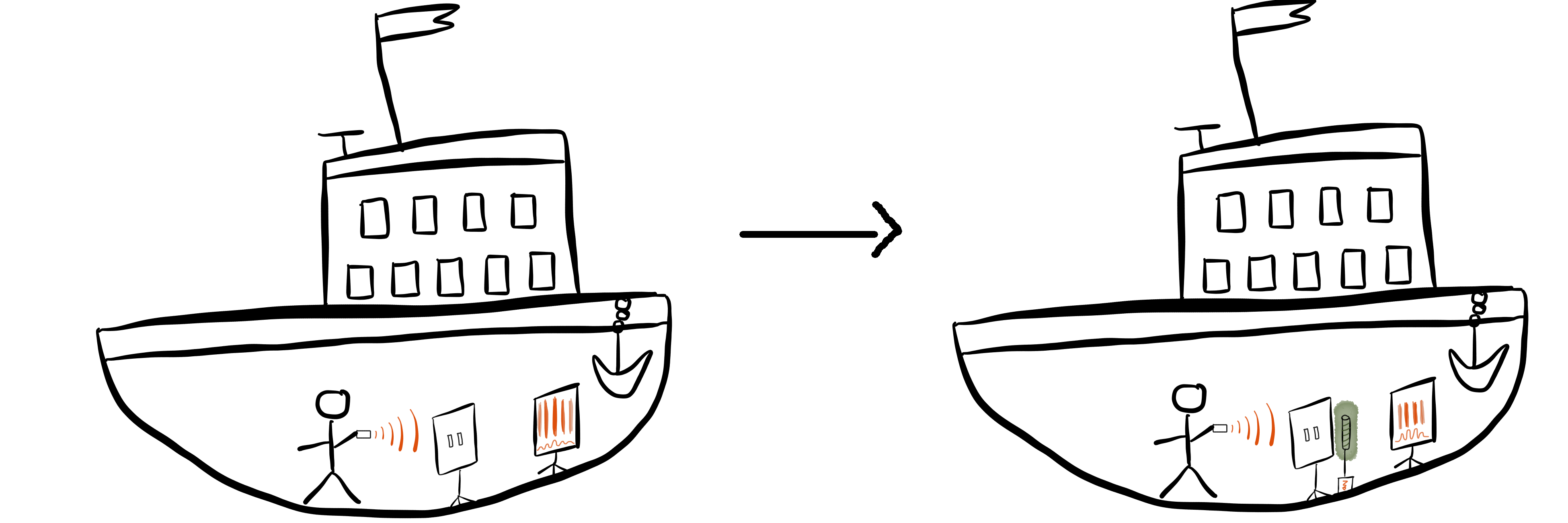}
\end{center}

Moreover, completely analogous to what we did in Eq.~\ref{eq:fieldstrengthtesnorcovariant} we can define the quantity\footnote{In our money toy model, this quantity encodes how much money we can money by trading money in a loop, i.e. about an arbitrage opportunity.}
\begin{equation} \label{eq:fieldstrengthtensoredyn}
    { F}_{\mu \nu}(x_\mu) \equiv  { \partial { A}_\nu \over \partial x^\mu } -  { \partial { A}_\mu \over \partial x^\nu  } \, ,
\end{equation}
which encodes in gauge invariant terms information about the presence of the electromagnetic field. In the context of Electrodynamics, the quantity in Eq.~\ref{eq:fieldstrengthtensoredyn} is known as the \textbf{field strength tensor}. 

With this in mind, we can now put the puzzle pieces together and disentangle real symmetries from redundancies in Quantum Mechanics and Electrodynamics.

\subsection{Putting the Puzzle Pieces Together}

First, we noted that there is a global symmetry in Quantum Mechanics (Eq.~\ref{eq:globalphaseshift})
\begin{equation} 
    \Psi(x) \to \mathrm{e}^{i\epsilon} \Psi(x) \, ,
\end{equation}
but not a local one (Eq.~\ref{eq:localphaseshift})
\begin{equation} 
    \Psi(x) \to \mathrm{e}^{i\epsilon(x)} \Psi(x) \, .
\end{equation}
We then learned that we can rewrite our equations such that they are invariant under local transformations by introducing bookkeepers $A_\mu$.

Afterwards, we argued that the invariance under global transformations represents a real symmetry, while the invariance under local transformations is only a redundancy. Formulated differently, we have invariance under active global transformations and passive local transformations if we write formulate the theory appropriately. However, Quantum Mechanics is not invariant under active local phase shifts.

Analogously to what we did in the money toy model, we then argued that our bookkepers $A_\mu$ can also appear as real dynamical parts of the system and not only as purely mathematical bookkeepers. The theory which describes the dynamics of the bookkeepers is Electrodynamics. The bookkeepers $A_\mu$ then become what we call the electromagnetic potential. 

The crucial point is then that as soon as we have a system where the bookkeepers are no longer purely mathematical objects but physical parts of it, they can induce measurable changes. An important example is the Aharonov-Bohm effect, where a nonzero potential induces a phase shift in the wave function \cite{Aharonov:1959fk}.

What this implies immediately is that, in principle, it's \textit{possible} to cancel any local phase shift using an electromagnetic potential $A_\mu$. However, it's important to take note that we still do not have a local symmetry when an electromagnetic potential is present, for example, in the ship in which our quantum physicist detects local phase shifts. 

It is instructive to reformulate this point using the notions "active transformation" and "passive transformation", which were introduced in Section~\ref{sec:activepassive}.

A passive transformation is simply a change of the coordinate systems and therefore cannot lead to any physical change. All we achieve through a passive transformation is a different description of the same physical situation. Therefore, when we perform a passive transformation, we must be careful to keep our description consistent. This means in particular that whenever we perform a local passive transformation, we have to accompany 
\begin{equation}
    \Psi \to \mathrm{e}^{i\epsilon(x)} \Psi 
\end{equation}
always with
\begin{equation}
    A_\mu \to    A_\mu + \partial_\mu \epsilon(x) \, .
\end{equation}
When we perform a \textit{passive} transformation these two transformations always go hand in hand. 

In contrast, an active transformation means that a real physical change happens and therefore the physical situation doesn't need to remain unchanged. In particular, when we induce an active local phase shift 
\begin{equation}
    \Psi \to \mathrm{e}^{i\epsilon(x)} \Psi \, ,
\end{equation}
the corresponding transformation of $ A_\mu$ does not happen automatically. Otherwise we wouldn't be able to detect local phase shifts in experiments.

However, it \textit{is} possible to cancel the phase shift in $\Psi$ through $ A_\mu$. But this only happens when we actively prepare the system in such a way, for example, by using an Aharonov-Bohm type setup. In other words, the active shifts in $\Psi$ and $ A_\mu$ are two separate transformations which can happen independently. 

To summarize: Quantum Mechanics and Electrodynamics are invariant under active global gauge transformations. Hence, global gauge symmetry is a real symmetry. However, only our description is invariant under passive local transformations. Therefore, local gauge symmetry is misnomer and should be better called local gauge redundancy.

Now, after these discussions of gauge symmetries in intuitive and concrete physical terms, it's time to move on and discuss how we can define them mathematically.

\section{Gauge Symmetries Mathematically}
\label{sec:gaugesymmathematically}

First of all, symmetries are described mathematically using group theory. A group consists of all transformations which leave the given system invariant and a binary operation which allows us to combine transformations.\footnote{For further details, see, for example, Ref.~\cite{schwichtenberg2018physics}}

In our money toy model the global symmetry group is the dilation group which consists of all possible dilatations 
\begin{align}
    f = e^\epsilon \, , \qquad \text{with} \qquad \epsilon \in \mathbb{R}\, .
\end{align}
of the given currency. The mathematical name for this group is $GL^+(1,R)$, the one-dimensional real general linear group with positive determinant. 

In Quantum Mechanics the global symmetry group is $U(1)$ and consists of all possible phase shifts
    \begin{align}
    f = e^{i \epsilon }  \, , \qquad \text{with} \qquad \epsilon \in \mathbb{R} \, .
\end{align}
The difference between $GL^+(1,R)$ and $U(1)$ is the factor of $i$ in the exponent of the transformation operators.\footnote{As a result, $U(1)$ is compact while $GL^+(1,R)$ is not.}

An important idea was then to introduce bookkeepers $A_\mu$ in order to make the theory locally redundant. These bookeepers allow us to use arbitrary local coordinate systems. In mathematical terms, we then have a local gauge symmetry, which is really just a redundancy in our description. So concretely, after the introduction of the bookkeepers, we can perform local dilatations\footnote{To unclutter the notation, we restrict ourselves here to one spatial dimension.}
\begin{equation}
    f (x)= e^{\epsilon(x)}  \, , \qquad \text{with} \qquad \epsilon(x) \in C^\infty \, 
\end{equation}
since our bookkeepers adjust accordingly (Eq.~\ref{eq:gaugetrafofieldcovariant}). Analogously, in Electrodynamics we can then perform local phase shifts (Eq.~\ref{eq:localphaseshift})
\begin{equation} 
    f(x) = e^{i\epsilon(x)} \, , \qquad \text{with} \qquad \epsilon(x) \in C^\infty \, .
\end{equation}
The crucial point here is that now our transformations parameters $\epsilon$ are functions of the location $x$. In other words, we can now shift the prices or analogously the phase at each point in space $x$ by a different amount. Without the bookkeepers only global shifts were permitted, which means that the prices or analogously the phases everywhere got shifted by exactly the same amount. 

After the introduction of the bookkeepers, we have the freedom to perform independent $GL^+(1,R)$ transformations at each point in space. This means in our money toy model we have a copy of the \textbf{gauge group} $GL^+(1,R)$ at each point in space. Taken together these copies yield the \textbf{group of gauge transformations}. 

Analogously, in Quantum Mechanics our symmetry group is $U(1)$ and we have a copy of $U(1)$ above each point in space.\footnote{It's crucial to keep the two notions "gauge group" and "group of gauge transformations" separate. While the former is finite-dimensional, the latter is a lot more complicated and infinite-dimensional. This comes about since the group of gauge transformations is a group of smooth functions on spacetime that take values in the gauge group.} Since each Lie group can be understood as a (differentiable) manifold, the situation can be understood geometrically using principal fiber bundles. For example, $U(1)$ transformations (Eq.~\ref{eq:globalphaseshift}) are unit complex numbers which lie on the unit circle in the complex plane. Hence, geometrically we can imagine $U(1)$ as a circle and that there is a circle attached to each point in spacetime.

As argued above, an important aspect of gauge theories is that local gauge symmetry is not a real symmetry but only a redundancy which is part of our description. This statement can lead to a lot of confusion and it is therefore necessary to carefully define what is meant by a local gauge symmetry.

If we look at our equations of Electrodynamics from a purely mathematical point of view, we notice that all transformations of the form $f(x) = e^{i\epsilon(x)}$, where $\epsilon(x)$ is a ${C}^\infty$ function, leave our description of the model unchanged. However, from experiments we know that global transformations ($\epsilon(x)= \text{const.}$) are real symmetries, while the invariance under local transformations is only a feature of our redundant description.\footnote{Often the notion local gauge symmetry is used to describe any kind of transformation of the form $f = e^{i\epsilon(x)}$. Since this definition also includes the constant function $\epsilon(x)=a$, it implies that global gauge symmetry is part of the local gauge group. (See, for example, page 64 in Ref.~\cite{duncan2012the}.) In other words, when we use this broad definition of the word local gauge symmetry, we are mixing symmetries and redundancies. }  

To keep these two kind of transformations separate, we define the group of local gauge transformations ${G}_\star$ as the set of transformations where the functions that parameterize the gauge transformations $\epsilon_\star(x)$ are only non-zero within a \textit{finite} region. This is a useful definition, since only global gauge transformations are real symmetries and these do not vanish if we move to the boundary at infinity.\footnote{Global transformations can, of course, also be redundancies. However, since local gauge transformations are exclusively redundancies it makes sense to single them out. Afterward we can discuss the remaining transformations in detail.} In more technical terms, we require here that the ${C}^\infty$ functions $\epsilon_\star(x)$ are functions with compact support \cite{nair2005quantum,Strocchi:2015uaa}. 

\begin{center}
    \includegraphics[width=0.5\textwidth]{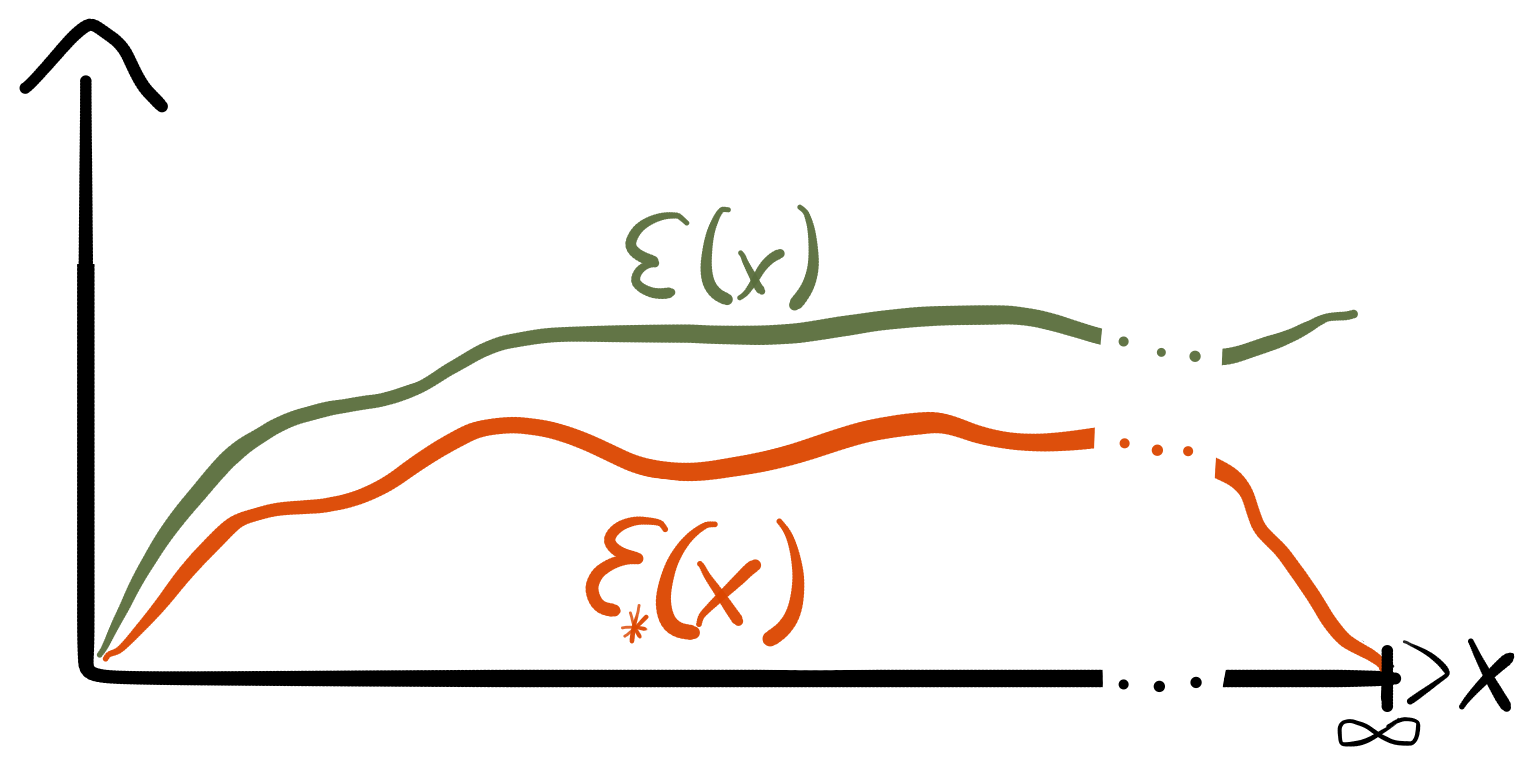}
\end{center}
As a result, the elements of ${G}_\star$ are only non-trivial within some finite region and in this sense completely localized
\begin{align} \label{eq:localgaugetrafodefmathematically}
{G}_\star &=  \big \{ \text{ set of all } f(x) \text{ such that }  f \to 1 \text{ as } |x| \to \infty \big \}  \, .
\end{align}
Now, to be able to define the real symmetries of the theory properly, we define the group ${G}_c$ as the set of all transformations which become constant at infinity, but not necessarily trivial 
\begin{align}
{G}_c &=  \big \{ \text{ set of all } f(x) \text{ such that }  f \to \text{ constant element of $G$, not necessarily $1$ as } |x| \to \infty \big \} \, .
\end{align}
This is a useful construction since now we can define the physical symmetry group as following quotient group
\begin{align} \label{eq:defglobalgaugegroup}
{G}_c / {G}_\star &\sim \text{ set of constant f's } \sim G \sim  \text{ physical global symmetry of the theory} 
\end{align}

There is another helpful way of looking at these definitions.

The first puzzle piece is that different boundary conditions correspond to physically distinct states of the system. In addition, different boundary conditions connected by a global symmetry correspond to states which are indistinguishable within the given subsystem. However, still we are dealing with different states since, in principle, it's possible to detect the changes made by bringing the subsystem in contact with the outside world. 

To understand this, let's consider once more Faraday's cage. A possible boundary condition for the electromagnetic potential is 
\begin{equation}
    A_\mu(x) \to 0 \quad \text{as} \quad |x| \to \infty \, .
\end{equation}
We can change the boundary conditions by charging the cage
\begin{equation}
    A_0(x) \to \phi_0 \quad \text{as} \quad |x| \to \infty \, .
\end{equation}
There is no way to detect this from the inside of the cage, but, in principle, it's possible to demonstrate that we changed the system by bringing the cage in contact with outside objects.

The crucial difference between redundancies and symmetries is that the states connected by the former are to be identified while the latter take us from one physical state to a different one. In particular this means that redundancies are not allowed to change the boundary conditions. The boundary conditions are preserved by transformations which become trivial at the boundary. This is exactly how we defined local gauge transformations in Eq.~\ref{eq:localgaugetrafodefmathematically}.

A final thing we need to talk about are our bookkeepers which we introduced to make the theory locally invariant. In mathematical terms the bookkeepers $A_\mu$ are called \textbf{connections}. A connection is a tool that allows us to compare prices or phases at different locations since it keeps track of how the local coordinate systems are defined and encodes information about the structure of the space we are moving in.

As already mentioned above, there are two situations where it's necessary to introduce connections. One the one hand, we need connections to allow arbitrary local coordinate systems. Here the connection keeps track of these local coordinate systems and lets us compare prices or phases defined according to different local conventions. On the other hand, connections are essential whenever the space we are interested in is curved.  

The prototypical example of a curved space is a sphere. To compare vectors at two different points on a sphere (e.g. to calculate a derivative), we need a procedure to move one vector to the location of the second one consistently.

The needed procedure is known as \textbf{parallel transport}. To understand it imagine we can imagine that we are walking on the sphere while holding a stick in your hand. While we are walking we our best to keep the stick straight. If we do this, we are parallel transporting the stick.

Mathematically, the infinitesimal parallel transport of a vector $V_\alpha(x)$ is defined as
\begin{equation}
    V_\alpha(x+dx) = V_\alpha(x) - \Gamma^{\alpha}_{\beta \gamma} (x) V^\beta(x) d x^ \gamma \, ,
\end{equation}
where $\Gamma^{i}_{jk}$ denotes the corresponding connection.

An important observation is that if the space we are moving in is curved, it's possible that the stick does not end up in its starting position if we move along a closed curve. 

\begin{center}
    \includegraphics[width=0.5\textwidth]{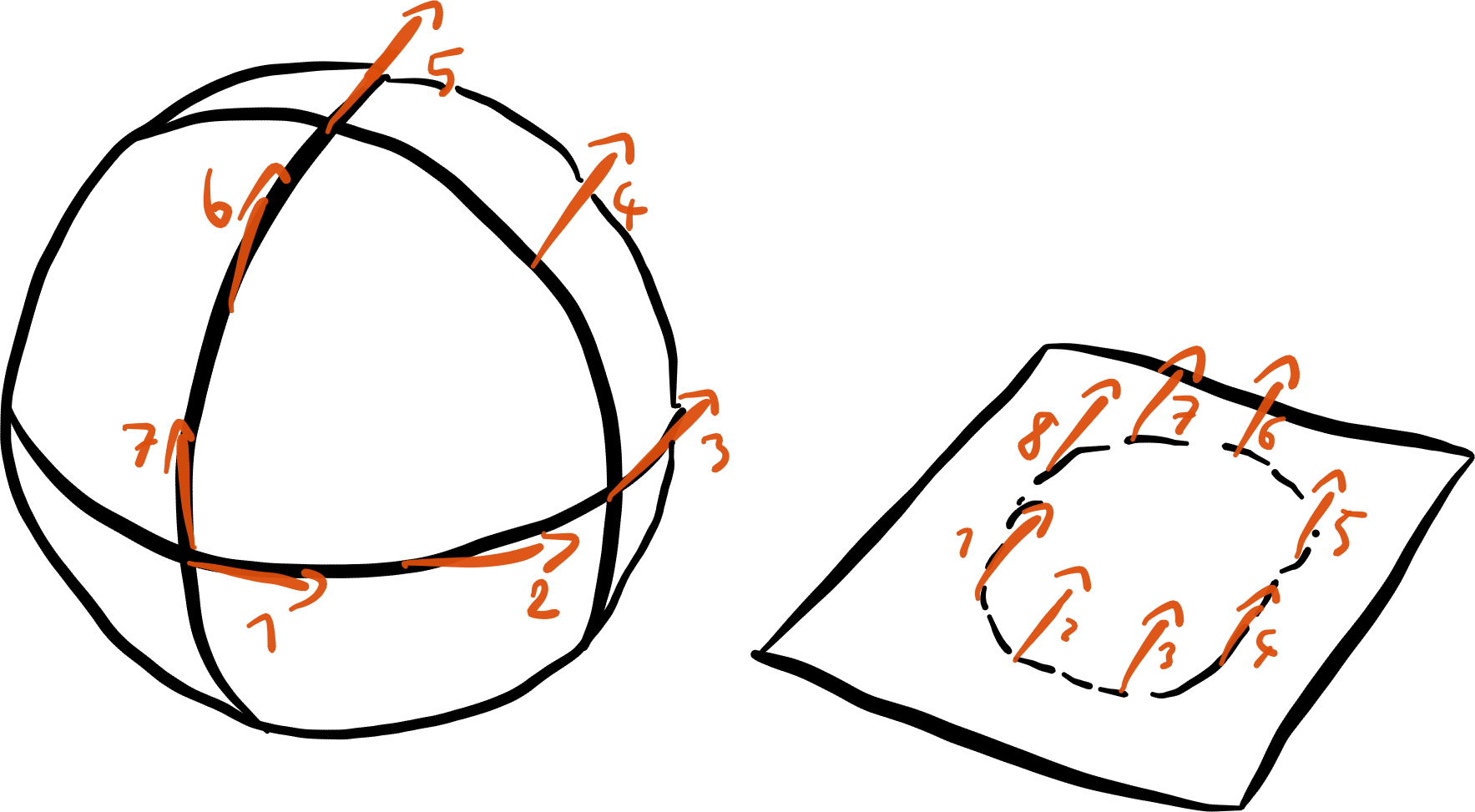}
\end{center}

Hence, the difference between the original vector and the vector which was parallel transported along an infinitesimal closed curve encodes information about the local curvature. Therefore, to define curvature we imagine that our vector moves from $A$ to $B$ via two different paths. 

\begin{center}
    \includegraphics[width=0.5\textwidth]{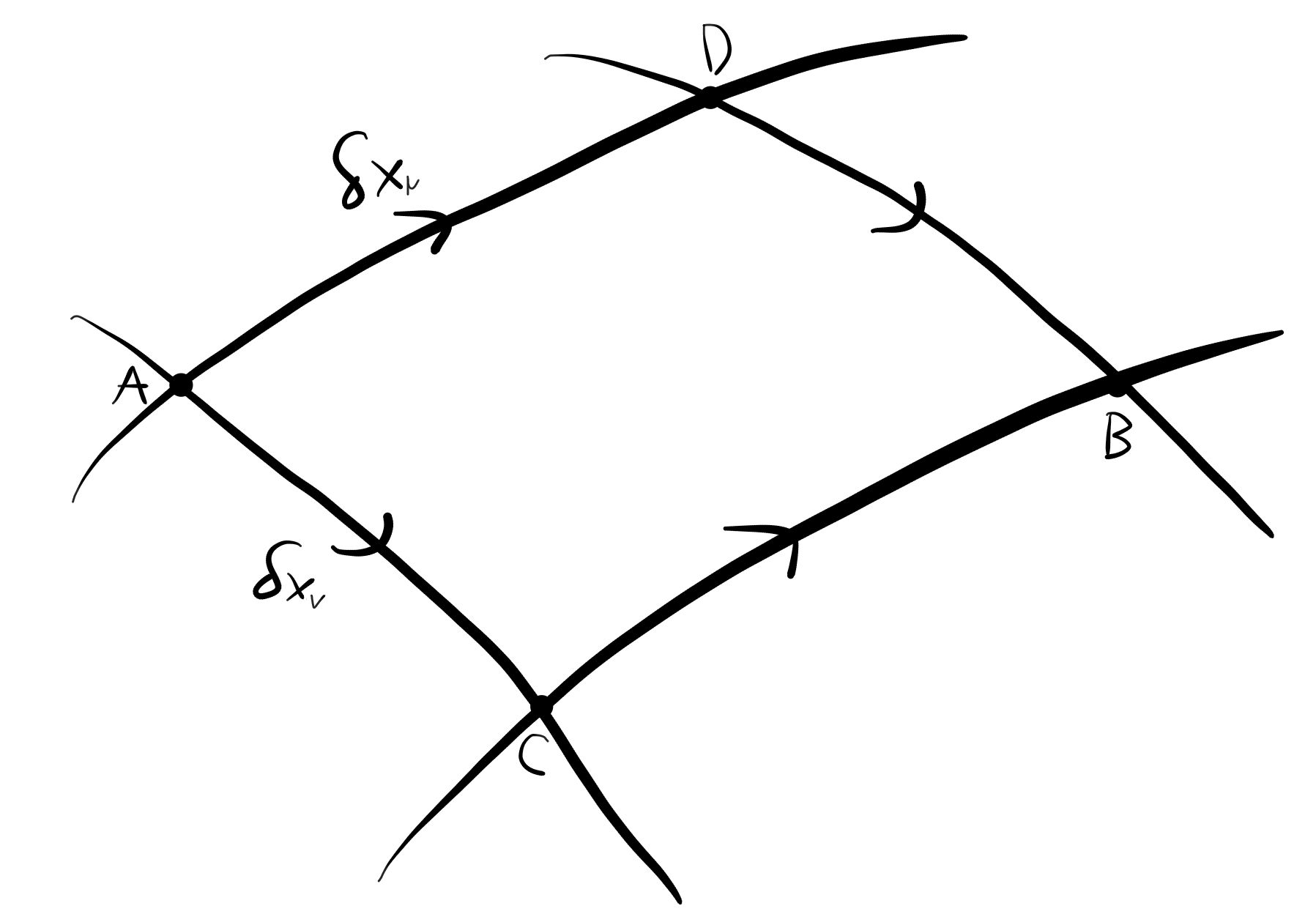}
\end{center}

Taken together these two paths yield a closed curve and we can calculate 
\begin{equation}
    V_\alpha(A\to C \to B) -  V_\alpha(A\to D \to B) = R_{\alpha\beta\;\mu}^{\;\;\;\nu} V^\beta dx^\mu dx^\nu + \ldots \, ,
\end{equation}
where $R_{\alpha\beta\;\mu}^{\;\;\;\nu}$ denotes the corresponding (Riemann) curvature tensor
        \begin{equation}
       R_{\alpha\beta\;\mu}^{\;\;\;\nu} = \partial_\alpha \Gamma_{\beta\;\nu}^{\;\,\mu} - \partial_\beta \Gamma_{\alpha\;\nu}^{\;\,\mu} + 
       \Gamma_{\alpha\;\kappa}^{\;\,\mu} \Gamma_{\beta\;\nu}^{\;\,\kappa} 
        - \Gamma_{\beta\;\kappa}^{\;\,\mu} \Gamma_{\alpha\;\nu}^{\;\,\kappa} \, .
       \end{equation}
In addition, it is possible that we have curvature in an internal space. For example, for the $U(1)$ symmetry described above, we can imagine that the various $U(1)$ copies are glued together non-trivially.

In this case, we again need a connection that allows us to move our wave function consistently from one point to another
\begin{equation}
   \Psi(x_\mu + \Delta x_\mu) = \Psi(x_\mu ) - {A}_\mu(x_\mu) \Psi(x_\mu )  \Delta x_\mu \, ,
\end{equation}
where ${A}_\mu(x_\mu)$ denotes the corresponding connection.
Moreover, completely analogously we can imagine that we don't end up with the same wave function when we move along a closed curve. If this is the case, we know that our internal space is curved and hence, we define
\begin{equation} 
    { F}_{\mu \nu}(x_\mu) \equiv  { \partial { A}_\nu \over \partial x^\mu } -  { \partial { A}_\mu \over \partial x^\nu  } 
\end{equation}
as a measure of the curvature. Take note that this is exactly how we defined in Eq.~\ref{eq:fieldstrengthtesnorcovariant} the quantity which encodes information about arbitrage opportunities and also in Eq.~\ref{eq:fieldstrengthtensoredyn} the quantity which tells us that there is a non-zero electromagnetic field. 

An important point is that connections can be nonzero, even though the curvature is zero. In this case, our connections are necessary only because of our choice of the local coordinate systems and not as a result of the physical situation itself. If the curvature is zero it is possible to find a choice of local coordinate systems such that no connection is necessary. However, whenever the curvature is nonzero, connections are essential and can't be removed by a clever choice of local coordinate systems. This will be discussed in detail in the next section.

There is, of course, a lot more that could be said about connections, curvature and fiber bundles. However, we will not go any further at this point, since we already have everything we need and there are already excellent introductions elsewhere.\footnote{For gentle introductions, see Ref.~\cite{Bernstein:1981xm} and Ref.~\cite{Marsh:2016hdj}.}

\section{Gauge Theory Mathematically}

\label{sec:gaugetheorymathematically}

The most important aspect of a gauge theory is that the connections get promoted from purely mathematical tools to real physical entities which behave according to their own equations of motion. While we can write any theory in such a way that arbitrary local coordinate systems are allowed, this does not transform the theory automatically into a gauge theory. Only when the connections are dynamical objects with their own equation of motion, we are dealing with a gauge theory. Otherwise, we have only written a theory in a more general but also more complicated way.

To understand this, let's consider concrete examples.

\subsection{Christoffel Symbols in Flat Spacetime}
        Christoffel symbols are usually associated with General Relativity and are necessary to take into account that spacetime can be curved. However, they also appear when we describe a given system in flat spacetime using curvlinear coordinates like, for example, spherical coordinates. 
        
        The components of a vector when we use rectangular coordinates remain unchanged in flat spacetime if we parallel transport it to another location. However, if we use spherical coordinates we need to take into account that the vectors' radial and tangential components do change \cite{ohanian2013gravitation}. In the picture below a vector (orange) is parallel transported from $A$ to $B$. The magnitude and direction of the vector are completely unchanged, but the radial (green) and tangential components (purple) do change. To take this consistently into account we need a connection.
        
        \begin{center}
    \includegraphics[width=0.5\textwidth]{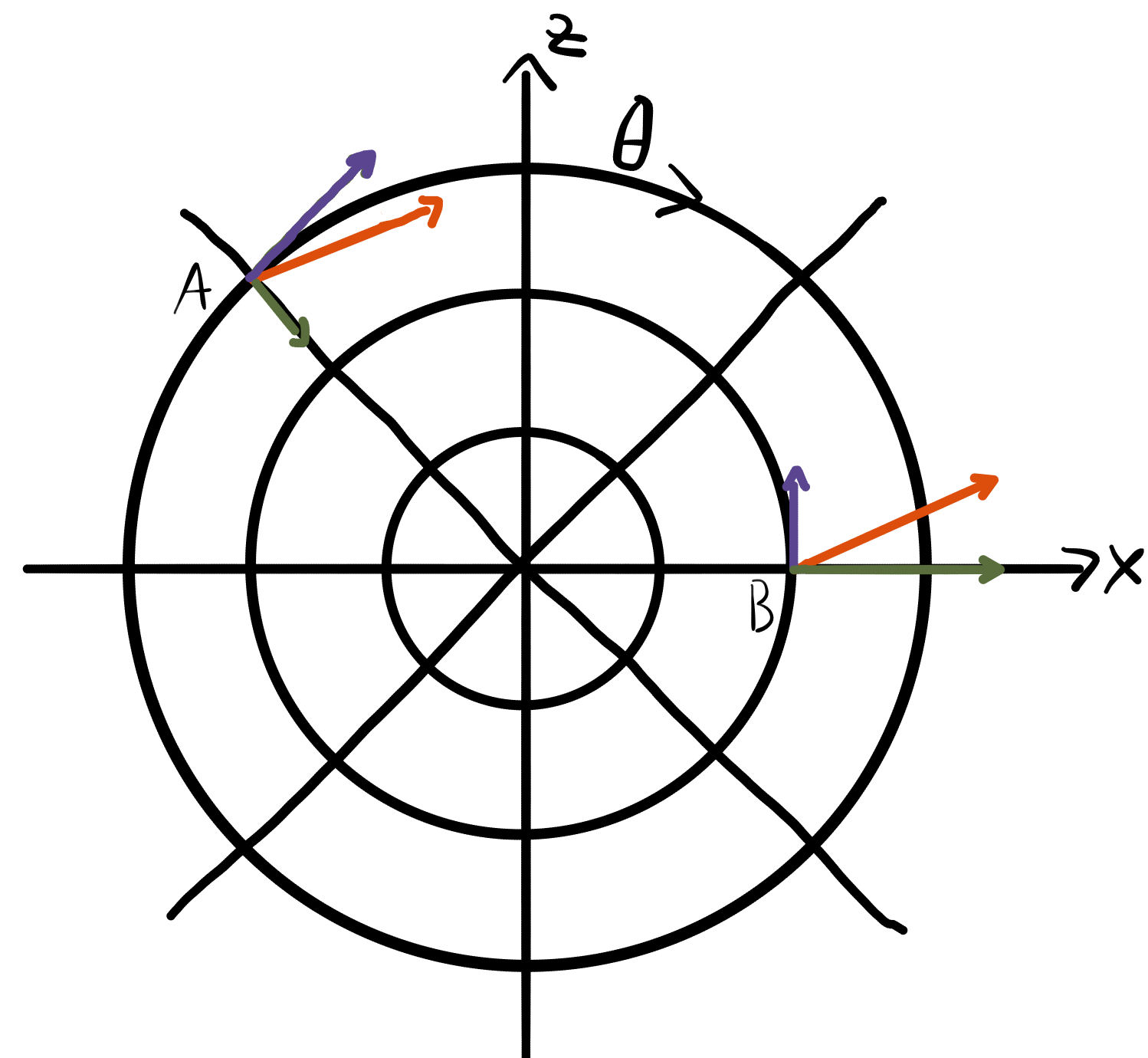}
\end{center}
        
        Concretely this means, for example, that if we rewrite the laws of Special Relativity in such a way that they hold in any coordinate system, we also have to introduce Christoffel symbols $\Gamma_{\mu \nu}^\mu$ \cite{Norton:2003cx}.\footnote{This is a "\textit{dirty little secret that they never tell you in class}", as John Baez puts it \cite{baezusenet}.}
        
        For example, in the usual rectangular coordinates, the equation of motion for a free particle reads
        \begin{equation}
            {d^2 x^\beta \over ds^2}  =0 \, ,
        \end{equation}
        If we want equations that hold in arbitrary coordinate systems, we need to modify this equation as follows
        \begin{equation}
           {d^2 x^\beta \over ds^2} + \Gamma_{\mu \nu}^\beta{d x^\mu \over ds } {d x^\nu \over ds } =0 \, ,
        \end{equation}        
        where $\Gamma_{km}^i$ are the Christoffel symbols and\footnote{In rectangular coordinates we have $ds^2 = \eta_{\mu \nu} dx^\mu dx^\nu$, where $\eta_{\mu \nu }$ is the Minkowski metric. } 
        \begin{equation}
            ds^2 = g_{\mu \nu} dx^\mu  dx^\nu  \, .
        \end{equation}
        However, the resulting theory is, of course, still Special Relativity and not General Relativity since all we have done is to rewrite the equations in a more general way.\footnote{In the early years of General Relativity this led to a lot of confusion. Einstein argued that general covariance is what sets General Relativity apart from all other theories. (A theory is generally covariant if it takes the same mathematical form in all possible coordinate systems.) However, already in 1917 the German physicist Erich Kretschmann responded that any theory can be rewritten in a general covariant way \cite{doi:10.1002/andp.19183581602}. One year later, Einstein responded: "\textit{Although it is true that every empirical law can be put in a generally covariant form, yet the principle of relativity possesses a great heuristic power}" \cite{doi:10.1002/andp.19183600402}. } The crucial point is that here the Christoffel symbols are purely mathematical bookkeepers which keep track of the local changes of the coordinate system. 
        
        The crucial point is that the connection $\Gamma_{\mu \nu }^\beta$ is subject to the constraint
        \begin{equation} \label{eq:flattnessconditionGR}
         R_{\alpha\beta\;\mu}^{\;\;\;\nu} =0 \, ,
        \end{equation}
        where $R_{ijkl}$ denotes the Riemann-Christoffel curvature tensor
        \begin{equation}
       R_{\alpha\beta\;\mu}^{\;\;\;\nu} = \partial_\alpha \Gamma_{\beta\;\nu}^{\;\,\mu} - \partial_\beta \Gamma_{\alpha\;\nu}^{\;\,\mu} + 
       \Gamma_{\alpha\;\kappa}^{\;\,\mu} \Gamma_{\beta\;\nu}^{\;\,\kappa} 
        - \Gamma_{\beta\;\kappa}^{\;\,\mu} \Gamma_{\alpha\;\nu}^{\;\,\kappa} \, .
       \end{equation}
        This constraint makes sure that we are always dealing with a situation which is physically equivalent to the situation that we started with. The only thing that is allowed to change is our description. In physical terms, Eq.~\ref{eq:flattnessconditionGR} tells us that we are dealing with a flat spacetime. As long as this condition holds, we know that it is possible to use coordinate transformations such that the connection vanishes everywhere, which is the exactly the situation we started with. The connection $\Gamma_{\mu \nu }^\beta$ is only indispensable when $R_{\alpha\beta\;\mu}^{\;\;\;\nu} \neq 0$, which means that spacetime is curved. If this is the case, it is not possible to get rid of $\Gamma_{\mu \nu }^\beta$ by a change of coordinate systems. Otherwise, $R_{\alpha\beta\;\mu}^{\;\;\;\nu}$ would also vanish, which corresponds to a different physical situation.
        
        In addition, in General Relativity the curvature is, in general, nonzero, but in addition, the Christoffel symbols are \textit{dynamical agents} with their own equation of motion (the Einstein equation). Only in this case, our bookkeepers become the dynamical actor which we call the gravitational field. (Otherwise, we are dealing with Special Relativity in curved spacetime.)

\subsection{Gauge Connections in Quantum Mechanics and the Money Toy Model}

Similarly, we can modify our equations in Quantum Mechanics to make them invariant under local $U(1)$ transformations. To achieve this, we need to introduce the connection ${ A}_\mu$. However, this connection is again a purely mathematical bookkeeper, as long as the curvature tensor vanishes\footnote{Here curvature is not referring to a property of spacetime but of our internal space.}
\begin{equation} \label{eq:flattnessconditionQM}
     { F}_{\mu \nu}(x_\mu) =0 \, ,
\end{equation}
where (Eq.~\ref{eq:fieldstrengthtesnorcovariantconti})
\begin{equation} \label{}
    { F}_{\mu \nu}(x_\mu) \equiv  { \partial { A}_\nu \over \partial x^\mu } -  { \partial { A}_\mu \over \partial x^\nu  } \, .
\end{equation}
 Moreover, completely analogous to what we discussed for Special and General Relativity above, ${ A}_\mu$ is only indispensable when ${ F}_{\mu \nu} \neq 0$. If $F_{\mu \nu}\neq 0$, we can't get rid of ${ A}_\mu$ at the same time everywhere by a change of coordinate systems, since this would imply ${ F}_{\mu \nu}=0$, which corresponds to a different physical situation. Moreover, only when there are nontrivial equations of motion for the connection (Maxwell's equations) it becomes the dynamical actor that we call the electromagnetic field.

Finally, we can also come back to our finance toy model. Here, we can introduce arbitrary local currencies and this makes it necessary to introduce bookkeepers $A_\mu(\vec n )$ which are able to handle the exchange of currencies. However, these bookkeepers are purely mathematical parts of our description as long as there is no arbitrage opportunity
\begin{equation}
    F_{\mu \nu}(\vec n) =0
\end{equation}
where (Eq.~\ref{eq:fieldstrengthtesnorcovariant})
\begin{equation} 
    F_{\mu \nu}(\vec n) = A_\mu(\vec n + \vec e_\mu) - A_\nu(\vec n ) -  [ A_\mu(\vec n + \vec e_\nu) - A_\mu(\vec n) ] \, ,
\end{equation}
If $F_{\mu \nu}(\vec n)=0$ it's possible to define a global currency such that $A_\nu(\vec n ) =0$ everywhere. The bookkeepers are only indispensable when there is at least one arbitrage opportunity $ F_{\mu \nu}(\vec n) \neq 0 $.  Moreover, as soon as there are equations of motion for the connections, they become dynamical actors.

So in summary, while connections also appear when we write down the equations of a given model in a more general way, they have no measurable effect since we are still describing the same model. The physical situation is only then a different one when the corresponding curvature is non-zero. In this case, our connections are no longer optional but essential parts of the model and have a measurable effect on the dynamics. Moreover, they become dynamical actors only when they change dynamically, i.e. they follow their own equations of motion.

\newpage

\section{The Gauge Tale}
\label{sec:gaugetale}

While local gauge symmetry is "only" a redundancy, it's certainly a useful concept. Rewriting our description of a given system in a redundant system makes the formulation more general and allows us to make suitable choices depending on the specific question we are trying to answer. In addition, local gauge symmetry can be useful, since it allows us, for example, to identify objects which are independent of local conventions. In our toy model we used this idea to identify the gauge invariant quantities $J_\mu$ and $F_{\mu \nu}$. 

Another important application of local gauge symmetry is as a didactic tool. In many Quantum Field Theory textbooks a variant of the following "gauge argument" is used to motivate the structure of interaction terms in the Lagrangian:\footnote{See, for example, \cite{sakurai1967advanced,schwichtenberg2018physics,aitchison2003gauge,mandl2010quantum,raifeartaigh1997the,ramond1990field,huang2010quantum}. Moreover, this is also how Yang and Mills motivated the structure of what is now known as Yang-Mills theories \cite{Yang:1954ek}.}

\begin{theo}[The Gauge Argument]
\begin{itemize}
    \item Our Lagrangian describing a free spin $1/2$ particle is invariant under global $U(1)$ transformations $\Psi \to \mathrm{e}^{i \Lambda} \Psi $.
    \item However, the Lagranian is not invariant under local $U(1)$ transformation $${\Psi \to \mathrm{e}^{i \Lambda(x)} \Psi }.$$ When we perform a local transformation, we get an additional term  $-(\partial_\mu \Lambda(x) \bar{\Psi} \gamma^\mu \Psi$ because of the product rule. This is puzzling since a global transformation "\textit{contradicts the letter and spirit of relativity, according to which there must be a minimum time delay equal to the time of light travel.}" \cite{ryder2003quantum} Or as David J. Gross puts it: "\textit{Today we believe that global symmetries are unnatural. They smell of action at a distance. We now suspect that all fundamental symmetries are local gauge symmetries".} \cite{Gross:1997bc} Therefore, we would expect that our theory is locally invariant and not just globally.
    \item There is a possible resolution of this puzzle, which starts by looking at the Lagrangian describing a massless free spin $1$ particle. It turns out that this Lagrangian is invariant under local $U(1)$ transformations $A_\mu \to A_\mu + \partial_\mu a(x)$.\footnote{The Lagrangian is not invariant under $A_\mu \to A_\mu + b_\mu(x) $. We are only allowed to add the total derivative of an arbitrary function.}
    \item The crucial idea is now that we can make our free spin $1/2$ Lagrangian invariant by using our spin $1$ field. Concretely, we combine the free spin $1/2$ Lagrangian with the free spin $1$ Lagrangian and add a new term of the form $A_\mu \bar{\Psi} \gamma_\mu \Psi$. Under the simultaneous transformation of $\Psi$ and $A_\mu$ this new term leads to an additional term which cancels the problematic term that led us to the conclusion that our Lagrangian is not invariant. 
    \item The final Lagrangian consisting of the free spin $1/2$ Lagrangian, the free spin $1$ Lagrangian and the additional term $A_\mu \bar{\Psi} \gamma_\mu \Psi$ correctly describes Quantum Electrodynamics.
\end{itemize}
\end{theo}

So in short, motivated by Special Relativity we demand that our global symmetry becomes a local one and this leads us to the introduction of a new term in the Lagrangian. The crucial point is now that this new term is exactly the right term that describes electromagnetic interactions.

This is certainly a persuasive story. However, as should be clear from the previous sections, it's not quite true \cite{10.1086/341848}.

Passive transformations never have any real-world implications since they only represent changes in how we describe a system. Therefore, there is no problem with Special Relativity when we switch without any time delay to a description using curvlinear coordinates. Whatever happens on some distant star does not care about how we describe it. No information needs to travel to the distant star before we can start to use a new coordinate system and therefore, there is no problem with Special Relativity.

So the argument to demand local invariance instead of the existing global symmetry because of Special Relativity is certainly not a rigorous one. In some sense, what most textbook do at this point is what Donald Knuth likes to call the "\textit{technique of deliberately lying}" in order to help students learn certain ideas easier \cite{knuth1986the}.\footnote{For example, Ryder writes: "\textit{the electromagnetic field arises naturally by demanding invariance of the action [...] under local (x-dependent) rotations}" \cite{ryder2003quantum}. This is, at least, not very precise. What really appears is the connection. But a connection is not necessarily a physical field. Only when the connection is also a dynamical actor it becomes field in the physical sense and the "gauge argument" does not offer a reason why this should happen.}

If we introduce connections solely to allow for arbitrary local coordinate systems, they aren't essential part of the description since we can get rid of them using local gauge transformations. So the reason why we introduce connections is by no means that they allow us to write down our equations in more general terms. 

The "gauge argument" does not lead us to a different physical physical system, but is only an argument about the description of a given system. We can see this immediately, since the curvature tensor is still flat after we have made our equations invariant under local transformations. So, as already mentioned above, there is no part of the "gauge argument" that tells us why the connections should be promoted from purely mathematical bookkeepers to real physical actors.

To summarize, the statement that we want locally invariant equations is a convenient way to motivate the introduction of connections. However, using the same argument we can introduce connections in any theory and certainly not every theory is a gauge theory. As argued above, the defining feature of gauge theories is that the connection is a dynamical part of the system.

\section{The Origin of Gauge Symmetry}
\label{sec:originofgaugesymmetry}

In the previous sections, we already discussed the origin of local gauge symmetry at great length. The main point is that local gauge symmetry is merely a redundancy that appears in our description when we want to allow arbitrary local coordinate systems. 

In contrast, global gauge symmetry is a real physical symmetry and its origin is far from trivial. Asking where global gauge symmetry comes from is completely analogous to asking where, for example, global rotational symmetry comes from. In some sense, wouldn't the absence of these symmetries be even stranger than their existence? Why shouldn't nature be rotational invariant? Rotational symmetry at the fundamental level is equivalent to the assumption that space is isotropic. An anisotropic space would need an explanation while an isotropic space is arguably the most natural situation.\footnote{Take note that here we are not talking about isotropy of matter in the universe on large scales. Instead, we are talking about the isostropy of spacetime itself.} Analogously, we can argue that a globally gauge invariant internal space is what we would expect naively.\footnote{$U(1)$ transformations are rotations in the complex plane. Hence, we can ask: why shouldn't our internal charge space be isotropic?}

It is even possible to turn the original question around and ask, for example, why there is \textit{only} Poincare symmetry? (Poincare symmetry is fundamental symmetry of Special Relativity). The Poincare group is not the biggest kinematical group possible \cite{Bacry:1968zf} and in addition, we can imagine fundamental laws that respect the even larger conformal group\footnote{Maxwell's equations are, in fact, invariant under conformal transformations \cite{Codirla:1997mz} and it's an "old dream" that the laws of nature become conformally invariant at high energies \cite{coleman1985aspects}. It is amusing to note that by taking thoughts like this serious it would have even been possible, for example, to predict "\textit{the expansion of the universe twenty years before it was discovered observationally by Hubble.}"\cite{Dyson:1972sd}}.

To summarize, the existence of global symmetries is not a miracle that needs explanation but rather the most natural situation we can imagine, given the existence of spacetime and internal spaces. Then, of course, the question arises where our internal spaces and spacetime comes from and why they have the properties that they have. However, a proper discussion of such deep questions is far beyond the scope of this paper.

Before we move on, we should discuss a line of arguments which recently got somewhat popular to explain the origin of gauge symmetry.\footnote{See, for example, Ref.~\cite{Arkani-Hamed:2016rak} or page 268 in Ref.~\cite{zee2010quantum}.} 

\begin{theo}[The Little Group Argument]
A massless spin $1$ particle like a photon has $2$ physical degrees of freedom.\footnote{A photon can only be transversely polarized but not longitudinally. Since a massless particle always travels at the speed of light and has no rest frame, it can only have its spin point along its axis of motion.} However, usually we use a Lorentz-covariant notation and therefore use $4$-vectors to describe photons. A real $4$-vector has $4$ degrees of freedom and therefore, by using a Lorentz-covariant notation for, say, a photon, we introduce a redundancy. Concretely this means that the $4$ components of the $4$-vector are not independent and it is possible to mix them in specific ways without changing anything.\footnote{As a reminder: Whenever we can perform a transformation without changing anything, these transformations are either redundancies of our description or symmetries of the system.} The redundancy introduced in this way is exactly what we call gauge symmetry. In more technical terms, the point is that we use irreducible unitary representations of the Poincare group to describe particles, while physical particle states transform instead as representation of the corresponding little group \cite{10.2307/1968551,Bargmann211,weinberg2005the}.\footnote{The little group for massive particles is $SO(3)$, while for massless particles the little group is $E(2)$, which is the Euclidean group that consists of $SO(2)$ and two-dimensional translations.}

\end{theo}

To examine the validity of this argument, let's consider a simple analogy: a marble that rolls along a circle. Since the movement of the marble is confined to the circle there there is only one degree of freedom. Therefore, a natural description uses the angle $\varphi$ to specify the location of the marble. However, another valid description uses Cartesian coordinates $(x,y)$. This description is redundant since it uses two coordinates $(x,y)$ to describe the one physical degree of freedom. This means that we can mix the two components $(x,y)$ by a rotation without inducing any measurable change.


What becomes clear here is that our conventional description using $(x,y)$ contains a redundancy because the system is rotational symmetric. As mentioned above, we can get rid of this redundancy by using $\varphi$ instead.\footnote{For photons this would mean that we use spinor-helicity variables instead of polarization vectors.} However, the crucial point is that the cause for the redundancy is that the system \textit{is} indeed symmetric. This rotational symmetry does not originate in our choice of description but is a fundamental feature of it. 

Analogously, global gauge symmetry is not an artifact of our description but a real feature of the systems we are describing. Of course, it can be useful to use that we've identified a certain symmetry to get a possibly simpler description. By using exclusively objects which are singlets under the corresponding symmetry group, we can effectively remove the symmetry from our description. However, the symmetry still crucially shapes the dynamics of the system and all we have done is choosing a description which makes the symmetry manifest.

\section{Summary and Conclusions}

We started our discussion of gauge symmetries in Section~\ref{sec:gaugesymintuitively} using a simple finance toy model. The gauge symmetry here is that it's possible to dilate currencies freely. If we want to use more than one currency, it's necessary that we introduce bookkeepers which keep track of how the local currencies change. 

We then argued that these bookkeepers can also be more than purely mathematical objects. For example, if there are imperfections in the exchange rates of the bookkeepers (arbitrage opportunities), it's possible to earn money purely by exchanging currencies. This way bookkeepers can have a real effect on the dynamics of the system. In addition, it's even possible that bookkeepers adjust their exchange rates dynamically. This way the bookeepers get promoted further from something that provides a static background to real dynamical actors. The corresponding dynamical rules are known as Maxwell's equations.

Afterward, we discussed gauge symmetry in the context of Quantum Mechanics and Electrodynamics. Here the gauge symmetry is that we can multiply our wave function with an arbitrary phase factor and can shift the electromagnetic potential arbitrarily with real numbers. We then argued that these global gauge symmetries are real symmetries, completely analogous to, for example, a global rotational symmetry.

An important observation was then that it's possible to combine Quantum Mechanics and Electrodynamics in such a way that the resulting description is invariant under local gauge transformations, e.g., local phase shifts $\mathrm{e}^{i\epsilon(x)}$. However, we then argued that experiments tells us that neither quantum mechanical nor electrodynamical systems are invariant under local gauge transformations. Therefore, local gauge symmetry is to be understood in a passive sense and only represents a redundancy in our description.

These observations were then used in Section~\ref{sec:gaugesymmathematically} to define global and local gauge symmetries properly. Local gauge transformations have only a nontrivial effect within a finite region and are therefore pure redundancies since they do not affect the boundary conditions. In contrast, global gauge transformations don't become trivial asymptotically and therefore represent real symmetries. 

Formulated differently, the main result is that local gauge symmetry is only a feature of our description, while global gauge symmetry is a feature the systems we are describing.

In Section.~\ref{sec:gaugetheorymathematically} we then discussed what really sets gauge theories apart from other types of theories. The main point is that as long as the corresponding curvature vanishes, the connection is a purely mathematical object with no influence on the dynamics.\footnote{Recall that connections show up in any theory if we want to formulate it such that arbitrary coordinate systems can be used.} Only when the curvature is non-vanishing \textit{and} the connections behave according to their own equations of motion, we are dealing with a gauge theory. 

In Section~\ref{sec:gaugetale}, we then discussed the common "gauge argument".\footnote{As a reminder: The usual "gauge argument" is that gauge theories are successful because we must promote our global symmetries to local symmetries because of Special Relativity and this requires the introduction of new fields. These new fields are then exactly what we need to describe the fundamental interactions.} We argued that this argument does not hold up to closer scrutiny since promoting a global symmetry to a local symmetry does not lead to a new theory because the curvature is still vanishing. 

Finally, in Section~\ref{sec:originofgaugesymmetry} we discussed the origin of gauge symmetries. As already mentioned above, local gauge symmetries are only a feature of our description and show up whenever we want them to show up. In contrast, global gauge symmetries are real symmetries and understanding their origin is as difficult as understanding, for example, the origin of global rotational symmetry.

There are dozens of topics directly related to gauge symmetries that we didn't discuss and which are far too complex to give them proper justice here. Nevertheless, at least a few comments are in order.

A common point of confusion is the relationship between local gauge symmetries and Noether's theorems. However, as soon as the distinction between local and global gauge symmetries is properly defined (c.f. Eq.~\ref{eq:localgaugetrafodefmathematically}) the situation is straightforward. Only global gauge symmetries give rise, using Noether's first theorem, to non-vanishing Noether charges. This is a result of their non-trivial asymptotic behaviour. In contrast, local gauge transformations are the subject of Noether's second theorem and give rise to constraints \cite{Brading:2000hc}.

Another common point of confusion is the topic of spontaneous gauge symmetry breaking. The story routinely told to students is that the spontaneous breaking of a global symmetry inevitably gives rise to Goldstone bosons and that Higgs, Englert et~al. discovered a loophole by using a local symmetry instead.\footnote{For example, in Ref.~\cite{guidry1999gauge}: "\textit{The difference between the Goldstone and Higgs modes is simply that the spontaneous
symmetry breaking occurs in the presence of a local gauge symmetry for the
Higgs mode.}" Or in Ref.~\cite{frampton2008gauge}: \textit{For global gauge invariance, spontaneous symmetry breaking gives rise to massless scalar Nambu–Goldstone bosons. With local gauge invariance, these unwanted
particles are avoided, and some or all of the gauge particles acquire mass. } Or in Ref.~\cite{quigg2013gauge}: "\textit{However, if the spontaneously broken symmetry is a local gauge symmetry, a miraculous interplay between the would-be Nambu– Goldstone boson and the normally massless gauge bosons endows the gauge bosons with mass and removes the Nambu–Goldstone boson from the spectrum. The Higgs mechanism, by which this interplay occurs, is a central ingredient in our current understanding of the gauge bosons of the weak interactions.}".} However, in all previous sections, we have argued that local symmetries are merely redundancies in our description. In addition, any theory can be rewritten in a locally invariant way but not every locally invariant theory is a gauge theory. Therefore, it is natural to wonder why local symmetries play such a crucial role in the Higgs mechanism. The confusion becomes even more pronounced through Elitzur's theorem \cite{Elitzur:1975im}, which in slogan form states that "\textit{a local gauge symmetry cannot break spontaneously}" \cite{wipf2013statistical}.\footnote{In some sense, local gauge symmetry is "\textit{too big to fail}". \cite{seiberttalk} } Moreover, even Englert himself emphasized in his Nobel lecture that "\textit{strictly speaking there is no spontaneous symmetry breaking of a local symmetry}" \cite{Englert:2014vga}. 

Therefore, it's clear that the breaking of a local symmetry is not really the loophole in Goldstone's theorem we are looking for. Instead, to understand why symmetry breaking in gauge theories is different, we have to focus on what really sets them apart. Arguably, the most important feature of gauge theories is the presence of long-range interactions. Global gauge symmetry forbids mass terms for the corresponding gauge bosons. Hence, without the Higgs mechanism the gauge bosons are massless and the corresponding interaction is long-ranged. When the Higgs field then gets a non-zero vacuum expecation value, the interactions become short-ranged and no Goldstone modes show up as a result of the previously long-ranged interactions. Formulated differently, the Goldstone degrees of freedom become the longitudinal degrees of the freedom of the now massive gauge bosons.\footnote{In this context, it is instructive to investigate the Higgs mechanism in completely gauge invariant terms \cite{Frohlich:1980gj,Frohlich:1981yi}. For a recent review of such methods, see Ref.~\cite{Maas:2017wzi}.}

Moreover, there are several interesting topics related to the constructions introduced in Section~\ref{sec:gaugesymmathematically} which, unfortunately, go far beyond the scope of this paper.

For example, what happens to all other possible gauge transformations, i.e. those that do not vanish and are non-constant at infinity?

Analogous to the construction in Eq.~\ref{eq:defglobalgaugegroup}, we can investigate ${G} / {G}_\star \, ,$ where ${G}$ is the group of \textit{all} allowed gauge transformations, i.e. not necessarily constant or trivial at infinity. The resulting group that we get through this construction is known as \textbf{asymptotic symmetry group} \cite{Strominger:2017zoo}. There are ongoing discussions about the interpretation of these asymptotic symmetries and they have been rediscovered several times \cite{Herdegen:2016bio,Gabai:2016kuf}.\footnote{For example, there is a deep connection to soft photon theorems and it's possible to argue that, in some sense, the asymptotic symmetries get broken spontaneously down to the usual non-angular dependent global symmetry since they aren't symmetries of the vacuum and the resulting Goldstone bosons are photons \cite{Ferrari:1971at,Avery:2015rga}. It's important to note that this idea is a different one from the \textit{speculative} idea that photons and gravitons are Goldstone bosons of spontaneously broken Lorentz invariance.} 

Another interesting observation is that a gauge transformations can be trivial at infinity: $ f(x) \to 1$ as $|x| \to \infty$, although $\epsilon(x)$ is non-zero. This is possible, because the function $\epsilon(x)$ appears in the exponent: $e^{i \epsilon(x)}$ and the exponential function is also equal to one for $\epsilon= 2 \pi$ or $ \epsilon = 4 \pi$ etc. The set of gauge transformations which is trivial at infinity, but for which the function that parametrizes the transformation is non-zero, are known as \textbf{large gauge transformations}. These large gauge transformations carry a non-zero winding number and they are important, for example, to understand the ground state of QCD \cite{Jackiw:1976pf,Callan:1976je}.

\bibliographystyle{h-physrev}
\bibliography{bib}

\begin{thebibliography}{10}

\bibitem{Redhead2003-REDTIO-3}
M.~Redhead,
\newblock The interpretation of gauge symmetry,
\newblock in {\em Symmetries in Physics: Philosophical Reflections}, edited by
  K.~A. Brading and E.~Castellani, pp. 124--139, Cambridge University Press,
  2003.

\bibitem{pittphilsci1831}
H.~Lyre,
\newblock Holism and structuralism in u(1) gauge theory, 2004.

\bibitem{pittphilsci9289}
A.~Afriat,
\newblock Shortening the gauge argument, 2012.

\bibitem{BELOT2003189}
G.~Belot,
\newblock Studies in History and Philosophy of Science Part B: Studies in
  History and Philosophy of Modern Physics {\bf 34}, 189  (2003).

\bibitem{Brading2004-BRAAGS}
K.~Brading and H.~R. Brown,
\newblock British Journal for the Philosophy of Science {\bf 55}, 645 (2004).

\bibitem{doi:10.1086/687936}
J.~O. Weatherall,
\newblock Philosophy of Science {\bf 83}, 1039 (2016).

\bibitem{Earman2001-EARGM-2}
J.~Earman,
\newblock Philosophy of Science {\bf 69}, 209 (2001).

\bibitem{healey2007gauging}
R.~Healey,
\newblock {\em Gauging what's real : the conceptual foundations of contemporary
  gauge theories} (Oxford University Press, Oxford New York, 2007).

\bibitem{Percacci:1984bq}
R.~Percacci,
\newblock {Role of Soldering in Gravity Theory},
\newblock in {\em {International Summer School on Elementary Particles and High
  Energy Physics: Differential Geometric Methods Shumen, Bulgaria, August
  16-19, 1984}}, 1984.

\bibitem{quigg2013gauge}
C.~Quigg,
\newblock {\em Gauge theories of the strong, weak, and electromagnetic
  interactions} (Princeton University Press, Princeton, New Jersey, 2013).

\bibitem{Gross:1997bc}
D.~J. Gross,
\newblock {The Triumph and limitations of quantum field theory},
\newblock in {\em {Conceptual foundations of quantum field theory. Proceedings,
  Symposium and Workshop, Boston, USA, March 1-3, 1996}}, pp. 56--67, 1997,
  hep-th/9704139.

\bibitem{maggiore2005a}
M.~Maggiore,
\newblock {\em A modern introduction to quantum field theory} (Oxford
  University Press, Oxford, 2005).

\bibitem{hamedtalk}
N.~Arkani-Hamed,
\newblock {Space-time, quantum mechanics and scattering amplitudes}, 2010.

\bibitem{schwartz2014quantum}
M.~Schwartz,
\newblock {\em Quantum field theory and the standard model} (Cambridge
  University Press, Cambridge, United Kingdom New York, 2014).

\bibitem{Tong:2016kpv}
D.~Tong,
\newblock {Lectures on the Quantum Hall Effect},
\newblock 2016, 1606.06687.

\bibitem{Nelson:1984gu}
P.~C. Nelson and L.~Alvarez-Gaume,
\newblock Commun. Math. Phys. {\bf 99}, 103 (1985).

\bibitem{zee2010quantum}
A.~Zee,
\newblock {\em Quantum field theory in a nutshell} (Princeton University Press,
  Princeton, N.J, 2010).

\bibitem{wen2004quantum}
X.~Wen,
\newblock {\em Quantum field theory of many-body systems : from the origin of
  sound to an origin of light and electrons} (Oxford University Press, Oxford
  New York, 2004).

\bibitem{Rovelli:2013fga}
C.~Rovelli,
\newblock Found. Phys. {\bf 44}, 91 (2014), 1308.5599.

\bibitem{Trautman:1970cy}
A.~Trautman,
\newblock Rept. Math. Phys. {\bf 1}, 29 (1970).

\bibitem{Wallace2014-GREECO}
D.~Wallace and H.~Greaves,
\newblock British Journal for the Philosophy of Science {\bf 65}, 59 (2014).

\bibitem{Ilinski:1997tj}
K.~Ilinski,
\newblock (1997), hep-th/9710148.

\bibitem{doi:10.1119/1.19139}
K.~Young,
\newblock American Journal of Physics {\bf 67}, 862 (1999).

\bibitem{Maldacena:2014uaa}
J.~Maldacena,
\newblock Eur. J. Phys. {\bf 37}, 015802 (2016), 1410.6753.

\bibitem{Aharonov:1959fk}
Y.~Aharonov and D.~Bohm,
\newblock Phys. Rev. {\bf 115}, 485 (1959),
\newblock [,95(1959)].

\bibitem{tHooft:1980aiu}
G.~'t~Hooft,
\newblock Sci. Am. {\bf 242N6}, 90 (1980),
\newblock [,78(1980)].

\bibitem{schwichtenberg2018physics}
J.~Schwichtenberg,
\newblock {\em Physics from Symmetry} (Springer, Cham, Switzerland, 2018).

\bibitem{duncan2012the}
A.~Duncan,
\newblock {\em The conceptual framework of quantum field theory} (Oxford
  University Press, Oxford, 2012).

\bibitem{nair2005quantum}
V.~P. Nair,
\newblock {\em Quantum field theory : a modern perspective} (Springer, New
  York, 2005).

\bibitem{Strocchi:2015uaa}
F.~Strocchi,
\newblock (2015), hist-ph/1502.06540.

\bibitem{Bernstein:1981xm}
H.~J. Bernstein and A.~V. Phillips,
\newblock Sci. Am. {\bf 245}, 94 (1981).

\bibitem{Marsh:2016hdj}
A.~Marsh,
\newblock (2016), math.DG/1607.03089.

\bibitem{ohanian2013gravitation}
H.~Ohanian,
\newblock {\em Gravitation and spacetime} (Cambridge University Press,
  Cambridge New York, 2013).

\bibitem{Norton:2003cx}
J.~D. Norton,
\newblock General covariance, gauge theories and the kretschmann objection.,
  2001.

\bibitem{baezusenet}
\url{https://groups.google.com/d/msg/sci.physics/hK--nwmcGV0/enlzgXZqG5kJ}.

\bibitem{doi:10.1002/andp.19183581602}
E.~Kretschmann,
\newblock Annalen der Physik {\bf 358}, 575.

\bibitem{doi:10.1002/andp.19183600402}
A.~Einstein,
\newblock Annalen der Physik {\bf 360}, 241.

\bibitem{sakurai1967advanced}
J.~J. Sakurai,
\newblock {\em Advanced quantum mechanics} (Addison-Wesley Pub. Co, Reading,
  Mass, 1967).

\bibitem{aitchison2003gauge}
I.~Aitchison,
\newblock {\em Gauge theories in particle physics : a practical introduction}
  (Institute of Physics Pub, Bristol Philadelphia, 2003).

\bibitem{mandl2010quantum}
F.~Mandl,
\newblock {\em Quantum field theory} (Wiley, Hoboken, N.J, 2010).

\bibitem{raifeartaigh1997the}
L.~Raifeartaigh,
\newblock {\em The dawning of gauge theory} (Princeton University Press,
  Princeton, N.J, 1997).

\bibitem{ramond1990field}
P.~Ramond,
\newblock {\em Field theory : a modern primer} (Westview Press, Boulder, Colo,
  1990).

\bibitem{huang2010quantum}
K.~Huang,
\newblock {\em Quantum field theory : from operators to path integrals}
  (Wiley-VCH, Weinheim, 2010).

\bibitem{Yang:1954ek}
C.-N. Yang and R.~L. Mills,
\newblock Phys. Rev. {\bf 96}, 191 (1954),
\newblock [,150(1954)].

\bibitem{ryder2003quantum}
L.~Ryder,
\newblock {\em Quantum field theory} (Cambridge Univ. Press, Cambridge, 2003).

\bibitem{10.1086/341848}
C.~A. Martin,
\newblock Philosophy of Science {\bf 69}, S221 (2002).

\bibitem{knuth1986the}
D.~Knuth,
\newblock {\em The METAFONTbook} (Addison-Wesley, Reading, Mass, 1986).

\bibitem{Bacry:1968zf}
H.~Bacry and J.~Levy-Leblond,
\newblock J. Math. Phys. {\bf 9}, 1605 (1968).

\bibitem{Codirla:1997mz}
C.~Codirla and H.~Osborn,
\newblock Annals Phys. {\bf 260}, 91 (1997), hep-th/9701064.

\bibitem{coleman1985aspects}
S.~Coleman,
\newblock {\em Aspects of symmetry : selected Erice lectures of Sidney Coleman}
  (Cambridge University Press, Cambridge Cambridgeshire New York, 1985).

\bibitem{Dyson:1972sd}
F.~J. Dyson,
\newblock Bull. Am. Math. Soc. {\bf 78}, 635 (1972),
\newblock [,347(1972)].

\bibitem{Arkani-Hamed:2016rak}
N.~Arkani-Hamed, L.~Rodina, and J.~Trnka,
\newblock Phys. Rev. Lett. {\bf 120}, 231602 (2018), 1612.02797.

\bibitem{10.2307/1968551}
E.~Wigner,
\newblock Annals of Mathematics {\bf 40}, 149 (1939).

\bibitem{Bargmann211}
V.~Bargmann and E.~P. Wigner,
\newblock Proceedings of the National Academy of Sciences {\bf 34}, 211 (1948).

\bibitem{weinberg2005the}
S.~Weinberg,
\newblock {\em The quantum theory of fields} (Cambridge University Press,
  Cambridge, 2005).

\bibitem{Brading:2000hc}
K.~Brading and H.~R. Brown,
\newblock (2000), hep-th/0009058.

\bibitem{guidry1999gauge}
M.~W. Guidry,
\newblock {\em Gauge field theories : an introduction with applications}
  (Wiley, New York Chichester, 1999).

\bibitem{frampton2008gauge}
P.~Frampton,
\newblock {\em Gauge field theories} (Wiley-VCH John Wiley distributor,
  Weinheim Chichester, 2008).

\bibitem{Elitzur:1975im}
S.~Elitzur,
\newblock Phys. Rev. {\bf D12}, 3978 (1975).

\bibitem{wipf2013statistical}
A.~Wipf,
\newblock {\em Statistical approach to quantum field theory : an introduction}
  (Springer, Berlin, 2013).

\bibitem{seiberttalk}
N.~Seibert,
\newblock {Duality and emergent gauge symmetry}, 2014.

\bibitem{Englert:2014vga}
F.~Englert,
\newblock Annalen Phys. {\bf 526}, 201 (2014).

\bibitem{Frohlich:1980gj}
J.~Frohlich, G.~Morchio, and F.~Strocchi,
\newblock Phys. Lett. {\bf 97B}, 249 (1980).

\bibitem{Frohlich:1981yi}
J.~Frohlich, G.~Morchio, and F.~Strocchi,
\newblock Nucl. Phys. {\bf B190}, 553 (1981).

\bibitem{Maas:2017wzi}
A.~Maas,
\newblock (2017), hep-ph/1712.04721.

\bibitem{Strominger:2017zoo}
A.~Strominger,
\newblock (2017), hep-th/1703.05448.

\bibitem{Herdegen:2016bio}
A.~Herdegen,
\newblock Lett. Math. Phys. {\bf 107}, 1439 (2017), 1604.04170.

\bibitem{Gabai:2016kuf}
B.~Gabai and A.~Sever,
\newblock JHEP {\bf 12}, 095 (2016), 1607.08599.

\bibitem{Ferrari:1971at}
R.~Ferrari and L.~E. Picasso,
\newblock Nucl. Phys. {\bf B31}, 316 (1971).

\bibitem{Avery:2015rga}
S.~G. Avery and B.~U.~W. Schwab,
\newblock JHEP {\bf 02}, 031 (2016), 1510.07038.

\bibitem{Jackiw:1976pf}
R.~Jackiw and C.~Rebbi,
\newblock Phys. Rev. Lett. {\bf 37}, 172 (1976),
\newblock [,353(1976)].

\bibitem{Callan:1976je}
C.~G. Callan, Jr., R.~F. Dashen, and D.~J. Gross,
\newblock Phys. Lett. {\bf B63}, 334 (1976),
\newblock [,357(1976)].

\end{thebibliography}

\end{document}